\documentclass{aa}  
\usepackage{graphicx}
\usepackage{txfonts}
\begin{document}
   \title{Elusive AGN in the {\it XMM-Newton} bright serendipitous 
survey\thanks{Based on 
observations collected at the Telescopio Nazionale Galileo (TNG) and
at the European
Southern Observatory (ESO), La Silla, Chile and on observations
obtained with {\it XMM-Newton}, an ESA science mission
with instruments and contributions directly funded by ESA Member States and 
the USA (NASA)}}

   \subtitle{}

   \author{A. Caccianiga\inst{1}
          \and
          P. Severgnini\inst{1}
	  \and
	  R. Della Ceca\inst{1}
	  \and
	  T. Maccacaro\inst{1}
          \and
	  F. J. Carrera\inst{2}
	  M. J. Page\inst{3}
	  %\fnmsep
          }

   \offprints{A. Caccianiga}

   \institute{INAF - Osservatorio Astronomico di Brera, via Brera 28, 
 I-20121 Milan, Italy\\
              \email{alessandro.caccianiga, paola.severgnini, roberto.dellaceca, tommaso.maccacaro@brera.inaf.it}
         \and
Instituto de F\'\i sica de Cantabria (CSIC-UC), Avenida de los
Castros, 39005 Santander, Spain\\
\email{carreraf@ifca.unican.es}
\and
           Mullard Space Science Laboratory, University College London, 
      Holmbury St. Mary, Dorking, Surrey, RH5 6NT\\
             \email{mjp@mssl.ucl.ac.uk}
             %\thanks{}
             }

   \date{}

% \abstract{}{}{}{}{} 
% 5 {} token are mandatory
 
  \abstract
   {Optical follow-up of X-ray selected sources finds a significant fraction of ``optically dull''
sources characterized by optical spectra without obvious signature of AGN activity. In many cases, however,
the presence of an AGN is inferred from other diagnostics (e.g. the X-ray properties). Understanding and 
accounting for this ``elusiveness'' is mandatory for a reliable study of the AGN physical and statistical 
properties.}
   {We investigate here the nature of all the sources (35 in total) in the XMM-Newton bright 
serendipitous survey 
(which is 86\% optically identified) 
showing an optical spectrum dominated by the light from the host galaxy with no evidence 
(or little evidence) for the presence of an AGN.}
   {We use the X-ray spectral analysis to assess the presence of an AGN in 
these sources and to characterize its properties.}
   {We detect AGN activity in 33 out of 35 sources. The remaining 2 sources 
are the ones with the lowest X-ray luminosity in the sample 
($L_{[2-10keV]}<$10$^{41}$ erg s$^{-1}$) and their X-ray emission could be 
produced within the host galaxy. We find that the ``recognition problem'' for AGN is very critical in
the low-luminosity regime (at least 60\% of the AGN with $L_{[2-10keV]}<$10$^{43}$ erg 
s$^{-1}$ are elusive) becoming negligible for high X-ray luminosities
($\sim$1.5\% of elusive AGN with $L_{[2-10keV]}>$10$^{44}$ erg s$^{-1}$). 
This problem affects mostly absorbed AGN ($\sim$40\% of type~2 AGN in the  
survey are elusive)
but also a significant fraction of unabsorbed AGN (8\%).
 }
{We find that
the simplest explanations of why these 33 (or most of them) AGNs are elusive are two: at 
low X-ray luminosities ($<$10$^{43}$ erg s$^{-1}$) 
the most important reason is the intrinsically low AGN/galaxy contrast (optical dilution) 
while at high luminosities ($>$10$^{44}$ erg s$^{-1}$) it is due to the optical 
absorption (in the Compton-thin regime, i.e. N$_H<$10$^{24}$ cm$^{-2}$).
Alternative hypotheses, like the presence of Compton-thick sources, BL Lac
objects or ``non-standard'' AGN (e.g. with $\alpha_{OX}<$1 or with weak/absorbed Narrow 
Line Region)  are not supported by the data  although
we cannot exclude the presence in the sample of a few sources of these types.}


   \maketitle

%__________________________________________________________________

\section{Introduction}
A great wealth of information on X-ray selected sources is usually
obtained from the analysis of their optical counterparts. Optical
spectroscopy, in particular, represents the fundamental step 
to determine not only the distance (for extragalactic objects) but also
a classification of the source. Whether the observed 
X-ray emission is attributed to a nuclear activity or not is often
inferred from the spectral 
properties of its optical counterpart. Also the characterization of the
``AGN type'' (type~1, type~2, BL Lac object, ...) is mostly based on
the information derived from the spectroscopic analysis. 
This ``optical step'', while often necessary due to the 
difficulty of obtaining this information directly from the X-ray 
data, 
can be potentially problematic when the optical strength of the
AGN, for whatever reason, is not expected to dominate on the 
host-galaxy emission. This ``recognition problem'' is obviously critical 
for intrinsically featureless AGN, like the BL Lac objects, as 
pointed out by Browne \& March\~a (1993), but it can be important also
for emission line AGN. Indeed, the presence of absorption or an intrinsic weakness of 
the AGN (or a combination of the two) may hide the AGN activity behind the
normal star-light of the host galaxy also in emission line sources. 
The optical elusiveness (or ``dullness'')
of some X-ray selected AGN is well known since the early surveys made with the 
{\it Einstein} Observatory and
has been recently pointed out as a critical problem when studying the nature
of the X-ray background and, 
in general, when studying the nature of X-ray selected sources 
(e.g. Elvis et al. 1981; Maccacaro et al. 1987; 
Griffiths et al. 1995; 
Comastri et al. 2002; Severgnini et al. 2003; Yuan \& Narayan 2004; 
Georgantopoulos \& Georgakakis 2005; Rigby et al. 2006).
Optical elusiveness is often associated to the existence of X-ray sources
with relatively large X-ray luminosity ($>$10$^{42}$ erg s$^{-1}$) and 
an early-type galaxy optical spectrum (the so-called XBONG, for X-ray Bright
Optically Normal Galaxies). As discussed by Maiolino et al. (2003), however, 
the correct definition of ``optically elusive AGN'' should include also sources
with starburst or LINER optical spectrum, coupled to an unusually (for these
types of objects) ``high'' X-ray luminosity and not restricted just to the
sources with an early-type galaxy spectrum. Indeed, ``hard'' (2-10 keV)  
X-ray observations have often revealed the presence of an ``hidden'' AGN in
sources optically classified as starburst galaxies (e.g. Della Ceca et al. 2001; 
Franceschini et al. 2003).

Apart from making the detection of nuclear activity difficult, a 
low AGN/host-galaxy relative strength in the optical band may also create
many problems for the correct definition of the AGN type. For instance, 
sources whose optical spectrum is characterized by an early-type galaxy
spectrum plus a broad H$\alpha$ line are 
generically classified as AGN, but establishing the amount of optical
absorption is not trivial: a broad H$\alpha$ line is observed both in 
``classical'' Seyfert~1, completely unaffected by optical absorption, 
and in Seyfert~1.8/1.9, where a significant level of absorption is expected
($A_V\sim$2-3 mag) and that are often grouped with Seyfert~2 in the
statistical analysis of AGN samples (e.g. Maiolino \& Rieke 1995). 

The impact of a systematic mis-classification of  elusive AGN on the
statistical analysis of a sample of X-ray selected AGN can be significant
as the percentage of these sources can be very high (40-60\% in  
deep surveys, e.g. Hornschemeier et al. 2001). 

In this paper we address this problem by analyzing 
in detail the properties of 35 X-ray sources 
selected in the {\it XMM-Newton} Bright Serendipitous survey (XBS,
Della Ceca et al. 2004) that show an optical spectrum which is dominated
by the host-galaxy light with no
or little evidence of the presence of an AGN. These 35 objects 
are selected from a statistically well defined survey with a high 
identification level ($\sim$86\%).
For this reason, once the nature of these sources has been established, 
we will be in the position of making firm statements on the statistical
importance of the optical recognition of AGN at different X-ray luminosities and
for different classes of objects.

After the definition  of the sample (Section~2) we 
use the X-ray data to assess the presence of an AGN in these 35 
objects (Section~3) and we study (Section~4) the possible reasons for their
optical elusiveness using a simple spectral model. 
We summarize the conclusions in Section~5. 
Throughout this paper H$_0$=65 km s$^{-1}$ Mpc$^{-1}$, $\Omega_{\Lambda}$=0.7 
and $\Omega_M$ = 0.3 are assumed.

 \section{Optically elusive AGN candidates in the XBS survey}

The {\it XMM-Newton} Bright Serendipitous Survey (XBS survey, 
Della Ceca et al. 2004)
is a wide-angle (28 sq. deg) high Galactic latitude ($|b|>$20 deg) 
survey based on the {\it XMM-Newton} 
archival data. It is composed by two samples both flux-limited 
($\sim$7$\times$10$^{-14}$ erg s$^{-1}$ cm$^{-2}$) in two separate energy 
bands: the  0.5-4.5 keV band (the BSS sample) and the 4.5-7.5 keV 
band (the ``hard'' HBSS sample). A total of 237 (211 for the HBSS sample) 
independent fields have been used to select   400 sources, 
389 belonging to the BSS sample and 67 to the HBSS sample (56 sources are
in common). The details on the fields selection strategy, the sources
selection criteria and the general properties of the 400 objects 
are discussed in Della Ceca et al. (2004). 

To date,  the spectroscopic identification  has been 
nearly completed 
and 86\% of the 400 sources have been already spectroscopicaly 
observed and classified. The results will be presented in Caccianiga et al.
(2007, in prep.).

%______________________________________________ Gamma_1 (lg rho, lg e)
   \begin{figure}
   \centering
    \includegraphics[width=9cm]{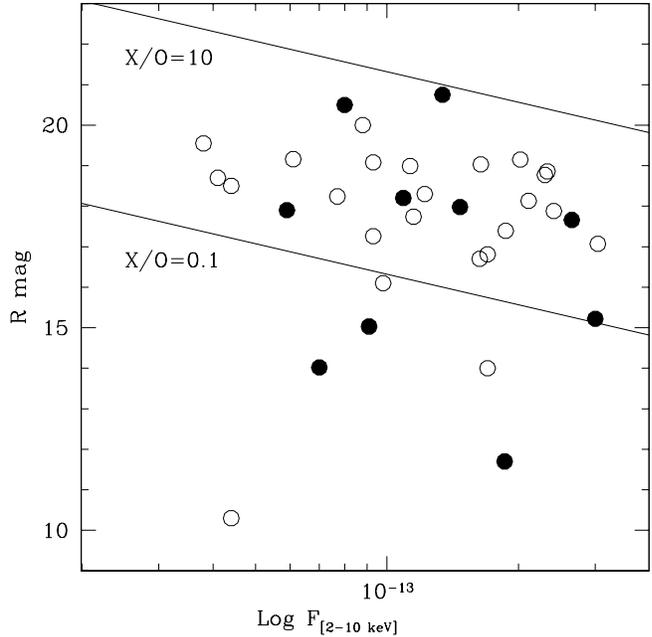}

   \caption{The magnitude/X-ray flux of the 35 elusive AGN candidates discussed in
this paper. Filled circles represent the objects with an early-type spectrum.) 
}
              \label{x_o}
    \end{figure}

A significant number of sources in the XBS survey  is characterized by the presence, 
in the optical spectrum, of a important contamination from the host galaxy star light: 
for instance in about 80 objects (i.e. $\sim$27\% of the extragalactic objects)
we detect a sharp 4000\AA\  break ($>$20\%),  a clear indication that the continuum is 
highly contaminated by the star-light from the host galaxy. In some cases, besides the
star-light continuum, we detect broad and/or narrow emission lines that allow 
to detect and correctly classify the X-ray source as AGN. But in a significant 
number of objects 
(20) the optical
spectrum does not offer a direct convincing evidence for the presence of an AGN.
These objects are either optically classified as galaxies with an
early-type spectrum (XBONG, 11 sources) or they can be classified as HII-region/starburst 
galaxies or LINERS on the basis of the diagnostic diagrams (e.g.  Veilleux \& Osterbrock 1987,
9 objects). 
We note that
a LINER spectrum does not necessarily imply the presence of an AGN (e.g. see discussion
in Maiolino et al. 2003).
 
In some other cases (10 objects), a broad 
($>$1000-2000 km/s) H$\alpha$ is possibly present 
but most of the remaining emission lines (in particular
the H$\beta$ emission line) are not detected. Even if the presence of 
an AGN in these sources
is suggested by the detection of a  broad H$\alpha$, we group also these
objects among the ``elusive'' AGNs. The reason is twofold: first, the correct
estimate of the H$\alpha$ line width is not trivial in these objects given the
relative weakness of the line and the presence of the two [NII] emission lines
that are often blended with the H$\alpha$. This means that the optical  evidence for an
AGN in these sources is often questionable. Second, this kind of object, if put at
larger redshift ($z>$0.3-0.4) and observed with the usual wavelength coverage 
(up to $\sim$8500-9000\AA) would appear as ``normal ellipticals'' and would be classified as 
XBONGs. Indeed, most of the sources classified as XBONGs in deep X-ray surveys (and 
4 out 11 XBONG in the XBS survey) do not
have optical spectra covering the H$\alpha$ region and, therefore, it is not possible
to apply a classification criterion based on the H$\alpha$ properties. For consistency with 
the literature it is therefore useful to classify also these sources as ``elusive'' AGN. 

Finally, 5 additional sources show the line 
[OIII]$\lambda$5007\AA\, which can be suggestive for the presence
of an AGN, but no H$\beta$ is detected, thus preventing 
a firm  classification of the source. Again, the classification as AGN and its
characterization is difficult also in these objects due to the host-galaxy light in
the optical spectrum.

We call  {\it optically elusive AGN 
candidates} the 35 sources (20+10+5) for which no-evidence or little evidence for the
presence of an AGN can be inferred from the optical spectrum. 
The reason why we believe that most of the sources in this group are actually
hiding an AGN is that the large majority of them 
have an (observed, i.e. not corrected for absorption) X-ray luminosity larger 
than 10$^{42}$ erg s$^{-1}$ cm$^{-2}$ and an
X-ray-to-optical flux ratio (using  the observed 2-10 keV flux and R 
magnitude) 
between 0.1 and 10 (see Fig.~\ref{x_o}), 
something which is strongly indicative of the presence of 
an AGN (Fiore et al. 2003). 

In Table~\ref{table} the optically elusive AGN candidates are listed 
together with a short description of their optical spectrum (column 4), the
redshift (column 5), the magnitude\footnote{Most of the magnitudes are taken from 
APM (Automated Plate Measuring machine facility {\it
www.ast.cam.ac.uk/$\sim$apmcat/}).
Since APM magnitudes of bright and extended sources can be systematically underestimated 
we have applied, in these cases, a correction discussed in March\~a et al. (2001)} (column 6) and 
the reference to the
optical spectrum (column~7). The optical spectra of the 30 objects observed in our own
spectroscopy plus an unpublished spectrum (XBSJ163332.3+570520) taken from the RIXOS project 
(Mason et al. 2000) are reported in Fig.~\ref{opt_spect}. More details on some of these
sources are discussed in Appendix~B.

\begin{table*}
\caption{Optical properties of the diluted AGN candidates}
\label{table}
\begin{tabular}{l c c l l r c}
\hline\hline
name & Sample & Optical position & spectral description & z & mag & reference \\ 
     &        &     (J2000)      &       &   &     &           \\
\hline
XBSJ000532.7+200716 & bs & 00 05 32.84 +20 07 17.4 & early type spectrum                      & 0.119 & 17.90 $^{(1)}$ & obs \\
XBSJ012540.2+015752 & bs & 01 25 40.36 +01 57 53.8 & narrow emission lines                    & 0.123 & 17.26 $^{(1)}$ & obs \\
XBSJ012654.3+191246 & bs & 01 26 54.45 +19 12 52.5 & early type spectrum                      & 0.043 & 13.70 $^{(2)}$ & obs \\
XBSJ013944.0--674909 & bs,hbs & 01 39 43.70 --67 49 08.1 & broad H$\alpha$                          & 0.104 & 17.74 $^{(1)}$ & obs \\
XBSJ014109.9--675639 & bs & 01 41 09.53 --67 56 38.7 & broad H$\alpha$                          & 0.226 & 19.15 $^{(1)}$ & obs \\
XBSJ014227.0+133453 & bs & 01 42 27.31 +13 34 53.1 & narrow emission lines                    & 0.275 & 19.08 $^{(1)}$ & obs \\
XBSJ021822.2--050615 & hbs & 02 18 22.16 --05 06 14.4 & early type spectrum                      & 0.044 & 15.22 $^{(2)}$ & obs \\
XBSJ025645.4+000031 & bs & 02 56 45.29 +00 00 33.2 & broad H$\alpha$?                         & 0.359 & 19.16 $^{(1)}$ & obs \\
XBSJ031859.2--441627 & bs,hbs & 03 18 59.46 --44 16 26.4 & broad H$\alpha$                          & 0.140 & 16.70 $^{(2)}$ & obs \\
XBSJ043448.3--775329 & bs & 04 34 47.78 --77 53 28.3 & early type spectrum                      & 0.097 & 17.66 $^{(1)}$ & obs \\
XBSJ050453.4--284532 & bs & 05 04 53.35 --28 45 31.0 & [OIII] and no H$\beta$                   & 0.204 & 18.99 $^{(2)}$ & obs \\
XBSJ051822.6+793208 & bs & 05 18 22.55 +79 32 09.8 & early type spectrum                      & 0.053 & 15.03 $^{(1)}$ & obs \\
XBSJ052116.2--252957 & bs & 05 21 16.08 --25 29 58.3 & broad H$\alpha$?                         & 0.332 & 19.55 $^{(2)}$ & obs \\
XBSJ052128.9--253032 & hbs & 05 21 29.04 --25 30 32.3 & early type spectrum                      & 0.588 & 20.75 $^{(2)}$ & obs \\
XBSJ075117.9+180856 & bs & 07 51 17.96 +18 08 56.0 & broad H$\alpha$                          & 0.255 & 19.03 $^{(1)}$ & obs \\
XBSJ083737.0+255151 & bs,hbs & 08 37 37.04 +25 51 51.6 & narrow emission lines                    & 0.105 & 17.07 $^{(1)}$ & obs \\
XBSJ090729.1+620824 & bs & 09 07 29.30 +62 08 27.0 & early type spectrum                      & 0.388 & 20.50 $^{(3)}$ & obs \\
XBSJ094526.2--085006 & bs & 09 45 26.25 --08 50 05.9 & broad H$\alpha$                          & 0.314 & 18.24 $^{(2)}$ & obs \\
XBSJ100032.5+553626 & bs & 10 00 32.29 +55 36 30.6 & narrow emission lines                    & 0.216 & 17.88 $^{(2)}$ &          1 \\
XBSJ101843.0+413515 & bs & 10 18 43.16 +41 35 16.5 & broad H$\alpha$?                         & 0.084 & 16.10 $^{(1)}$ & obs \\
XBSJ111654.8+180304 & bs & 11 16 54.72 +18 03 05.9 & narrow emission lines                    & 0.003 & 10.30 $^{(2)}$ &          2 \\
XBSJ112026.7+431520 & hbs & 11 20 26.62 +43 15 18.2 & broad H$\alpha$                          & 0.146 & 17.39 $^{(1)}$ & obs \\
XBSJ122017.5+752217 & bs & 12 20 17.76 +75 22 15.2 & early type spectrum                      & 0.006 & 11.70 $^{(2)}$ &          3 \\
XBSJ133942.6--315004 & bs,hbs & 13 39 42.47 --31 50 03.0 & broad H$\alpha$                          & 0.114 & 16.81 $^{(1)}$ & obs \\
XBSJ134656.7+580315 & hbs & 13 46 56.75 +58 03 15.7 & early type spectrum                      & 0.373 & 17.98 $^{(1)}$ & obs \\
XBSJ140100.0--110942 & bs & 14 00 59.93 --11 09 40.8 & narrow emission lines                    & 0.164 & 18.70 $^{(2)}$ & obs \\
XBSJ142741.8+423335 & hbs & 14 27 41.62 +42 33 38.1 & strong [OIII] and no H$\beta$            & 0.142 & 18.13 $^{(1)}$ & obs \\
XBSJ143835.1+642928 & bs,hbs & 14 38 34.72 +64 29 31.1 & narrow emission lines                    & 0.118 & 18.86 $^{(1)}$ & obs \\
XBSJ143911.2+640526 & hbs & 14 39 10.72 +64 05 28.9 & early type spectrum                      & 0.113 & 18.20 $^{(1)}$ & obs \\
XBSJ161820.7+124116 & hbs & 16 18 20.82 +12 41 15.4 & strong [OIII] and no H$\beta$            & 0.361 & 20.00 $^{(3)}$ & obs \\
XBSJ163332.3+570520 & bs & 16 33 31.94 +57 05 19.9 & [OIII] and no H$\beta$                   & 0.386 & 18.50 $^{(1)}$ &          4 \\
XBSJ193248.8--723355 & bs,hbs & 19 32 48.56 --72 33 53.0 & narrow emission lines                    & 0.287 & 18.77 $^{(1)}$ & obs \\
XBSJ230401.0+031519 & bs & 23 04 01.18 +03 15 18.5 & early type spectrum                      & 0.036 & 14.02 $^{(1)}$ & obs \\
XBSJ230434.1+122728 & bs & 23 04 34.25 +12 27 26.2 & strong [OIII] and no H$\beta$            & 0.232 & 18.30 $^{(1)}$ & obs \\
XBSJ231546.5--590313 & bs & 23 15 46.76 --59 03 14.5 & narrow emission lines                    & 0.045 & 14.00 $^{(2)}$ &          5 \\

\hline
\end{tabular}

column 1: name

column 2: sample (bs = [0.5-4.5 keV] BSS sample, hbs = [4.5-7.5 keV] HBSS sample)

column 3: position of the optical counterpart

column 4: a short description of the optical spectrum

column 5: redshift

column 6: red magnitude: (1) = derived from APM E-magnitude; (2) red magnitude (either R or r) 
from the literature or computed from our own photometry; 
(3) = magnitude visually estimated from the DSS red plates

column 7: references to the spectrum. 

obs = from our own spectroscopy; 

1 (SDSS J100032.24+553631.0) = spectrum from the SDSS (classified as galaxy). Also in RIXOS 
(Mason et al 2000) and in Zhou et al. (2006) 
(Classified respectively as Sy~2 and NLSy1, see Appendix~B for more details);
  
2 (NGC3607) = spectrum from Ho et al. 1995. Classified as LINER~2 in Ho et al. 1997;

3 (NGC4291) = spectrum from Ho et al. 1995;

4 (RIXOS F223 097) = In Mason et al. 2000 (the unpublished optical spectrum from RIXOS is 
reported in Fig.~\ref{opt_spect}); 

5 (IRAS 23128-5919) = spectrum from Kweley et al. 2001. In the literature this object 
(an ULIRG) is classified as Sy2/Starburst.

\end{table*}

%                                     two column figure (place early!)

   \begin{figure*}
   \centering
    \includegraphics[width=18cm]{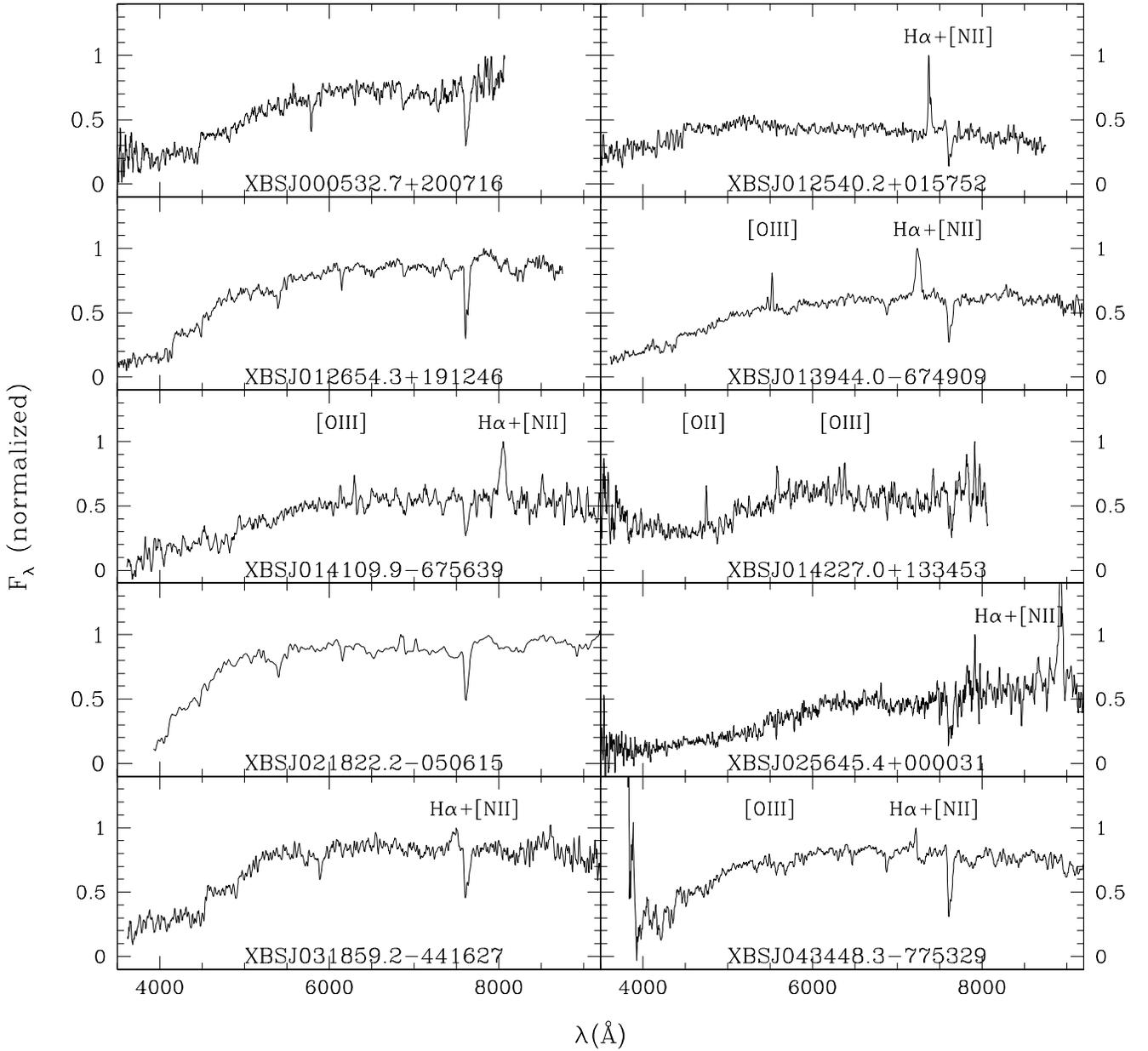}

   \caption{The optical spectra of the elusive AGN candidates (we do not report the 4 
spectra already published in the literature, see Tab.~\ref{table}). 
The strongest emission lines are marked.
}
              \label{opt_spect}%
    \end{figure*}

\addtocounter{figure}{-1}
%                
%                                     two column figure (place early!)

   \begin{figure*}
   \centering
    \includegraphics[width=18cm]{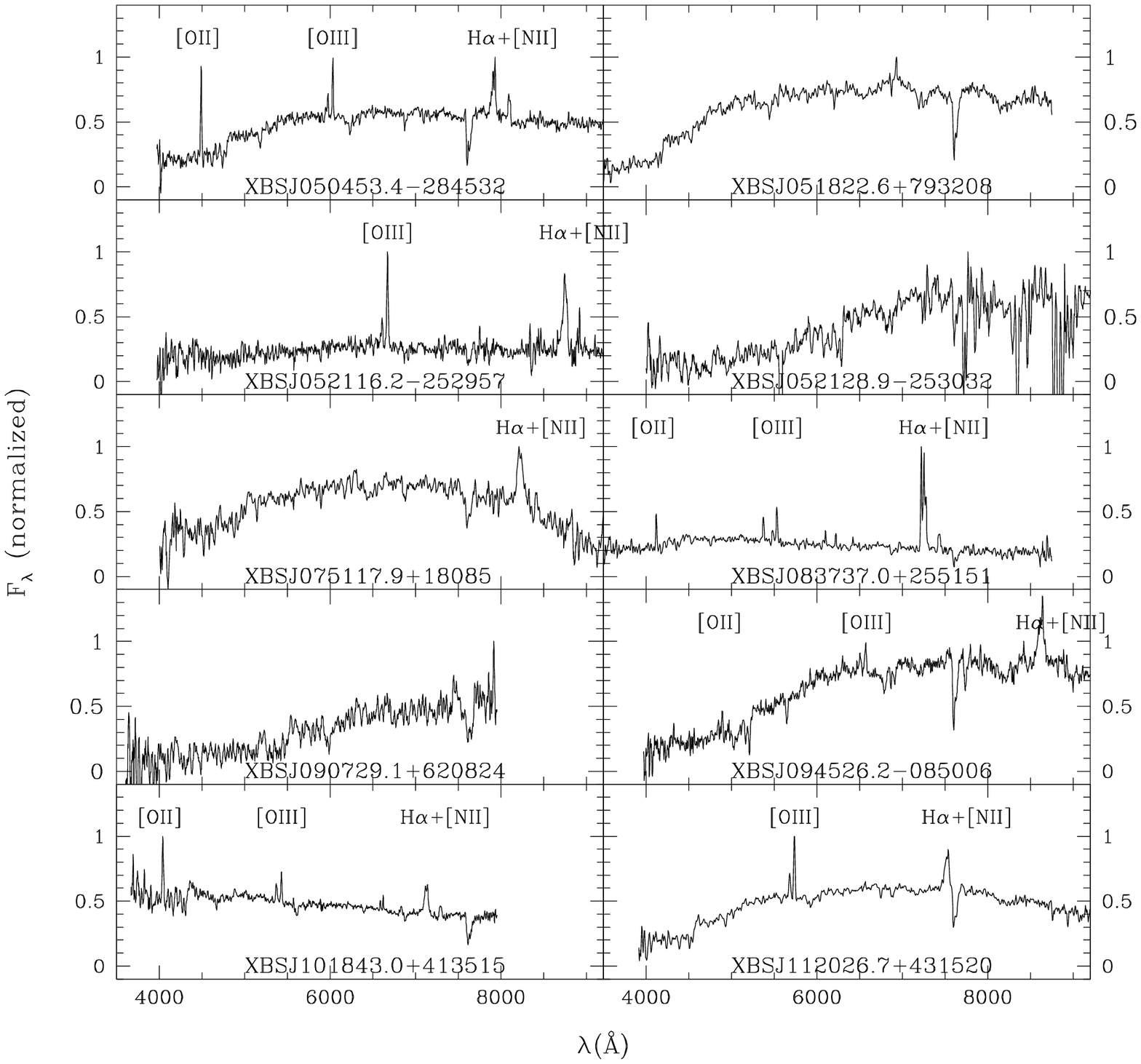}

   \caption{continue
}
           
    \end{figure*}
          \addtocounter{figure}{-1}
%                    two column figure (place early!)

   \begin{figure*}
   \centering
    \includegraphics[width=18cm]{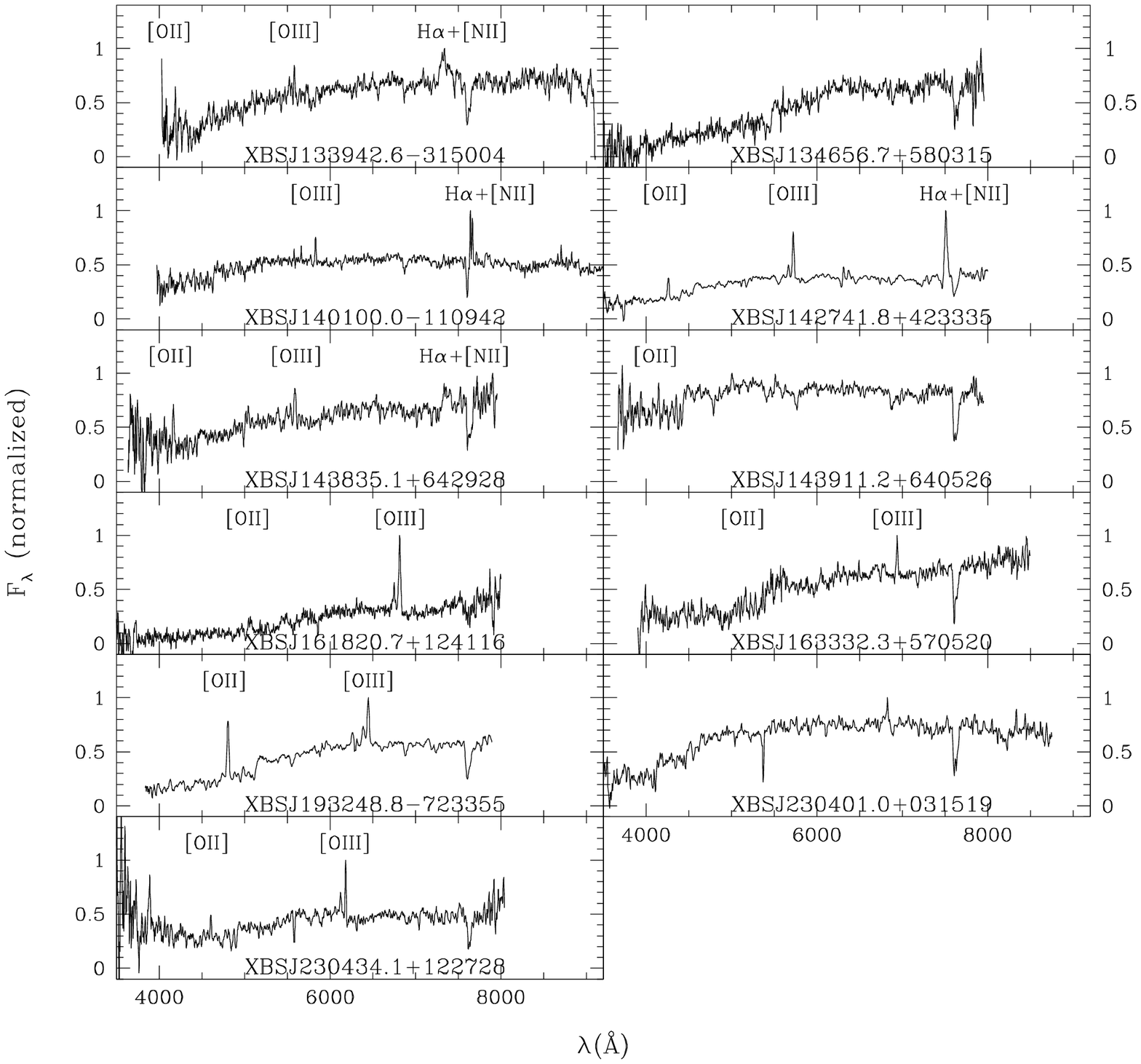}

   \caption{continue
}
             
    \end{figure*}

\section{X-ray spectral analysis}

Given the difficulty (or the impossibility) of finding  from the optical
spectra direct and strong 
evidence for the presence of an AGN  in the
35 sources discussed here, we use as starting point the X-ray data, that
are available, by definition, for the entire group of objects.
The X-ray data have the advantage with respect to the optical band of being 
less affected by the presence of non-nuclear emission so that the AGN/galaxy
contrast is expected to be larger. This is particularly true when 
the X-ray data cover hard energies ($>$2 keV), as in the case of 
{\it XMM-Newton} data. 

For 4 sources (XBSJ021822.2$-$050615, XBSJ031859.2$-$441627, 
XBSJ075117.9$+$180856 and 
XBSJ231546.5$-$590313) a {\it XMM-Newton} analysis has been already
presented and discussed in Severgnini et al. (2003) (the former three
sources) or in Franceschini et al. (2003) (the latter one) and, therefore,
we keep the results from that analysis and report the numbers published
in these two papers. An additional source (XBSJ012654.3$+$191246), a radio-loud
source, has been already presented in Galbiati et al. (2005) but we have re-done
the X-ray analysis using a different model.

For the X-ray spectral analysis of the remaining 31 sources we have 
considered  all the epic detectors available for each source. In general, data
from the 2 MOS cameras are always available, while the epic-pn
is missing in 8 cases (detector not available or sources outside the
field-of-view or under a CCD gap). 

Before the extraction of the spectra we have removed the time
intervals with a high background. The source extraction region 
is a circular one with a radius of 20$\arcsec$-30$\arcsec$,
depending on the source off-axis distance. The background is extracted
in a nearby source-free circular region of radius $\sim$2 times larger
than the one used for the source. The data from the two MOS cameras
have been combined after the extraction if the same blocking filter 
is used. MOS and pn data have been re-binned in order to have at
least 15-20 counts per channel, depending on the brightness of the
source.

We have used XSPEC (ver. 11.3.1) to perform a simultaneous 
fitting analysis of the MOS and pn data leaving the relative normalization
free to vary. All the relevant quantities (fluxes, luminosities, ...)
are computed using the MOS normalization (or MOS2 if the two MOS cameras
are not combined together). 

As usual, errors are given at the 90\% confidence level for one interesting
parameter ($\Delta\chi^2$=2.71).

We have always tried to fit the spectra both with a simple
absorbed power-law model and with a thermal model (Mekal). 
The reason for 
choosing these two models is that we expect most of the  sources analysed here
to contain an AGN (either absorbed or unabsorbed) or, alternatively,
that the observed X-ray emission is due to the galaxy or to a
group/cluster of galaxies. 

{\bf Power-law model.}
In the large majority of cases (27/35) a simple absorbed power-law model
offers an acceptable description of the data. In 3 additional cases 
this model is unable to give a good fit, leaving significant
residuals in the highest energy bins. In these cases also the thermal
model is rejected. We find that a better fit is obtained with 
a more complex model, composed by two  power-laws, one absorbed and one 
unabsorbed,
having the same spectral index ({\it leaky absorbed power-law model}). 
This model has different physical interpretations. For instance, the
leaky-model can represent the physical case where the X-ray primary source is 
observed both directly, through a cold absorbing medium, and scattered 
(not absorbed) 
by a warm, highly ionized gas located outside the absorbing medium. 
The {\it leaky absorbed power-law model} has been used by Franceschini 
et al. (2003) also in the case of the source XBSJ231546.5$-$590313 
(IRAS~23128$-$5919) with the addition of a thermal component (kT$\sim$0.65 keV).
 
In 4 other  cases a thermal model (black-body
or Mekal model) is required in addition to the 
power-law component, to better describe the low-energy
part of the spectrum. 

When the power-law model  is statistically
acceptable but the counts are too few ($<$100 counts) to well constrain 
both the absorption and the spectral index, we have decided to fix the
spectral index to $\Gamma$=1.9, which is the mean value observed in the
XBS survey of AGN (see Caccianiga et al. 2004 and Della Ceca et al. 2007 in 
prep),  to 
better constrain the absorption. 
In one further case (XBSJ142741.8+423335) we have decided to fix the value
of $\Gamma$ to 1.9 since the best-fit $\Gamma$ is much 
flatter (1.1) than the flattest $\Gamma$ observed among the unabsorbed AGN of the XBS survey 
($\sim$1.5) and we suspect that the combination absorption/relatively low-statistics 
leads to an underestimate of both $\Gamma$ and N$_H$. By fixing the value
of $\Gamma$ to 1.9 we find an absorption which is $\sim$70\% higher than the
one estimated by leaving $\Gamma$ free to vary. In both cases, however, the presence of
a relatively large absorption ($>$2$\times$10$^{22}$ cm$^{-2}$) is confirmed, 
i.e.
the classification of the source as ``absorbed AGN'' (see below) does not depend on the 
assumption on $\Gamma$. 

{\bf Thermal model.}
As a first step, we consider acceptable a thermal model (Mekal) which gives a
statistically significant fit ($\chi^2$ probability larger than 5\%)
with values of a temperature below 10 keV.  Usually 
the thermal model is statistically rejected for  ``flat'' spectra, i.e. 
those well fitted by flat ($\Gamma<$1.7) and/or absorbed power-laws. 
Then we have further analysed the cases where a thermal model is statistically
acceptable to decide whether the best-fit temperature is physical or not,
considered the X-ray luminosity of the source. It is well known, in fact,
that in both clusters of galaxies and ``normal'' elliptical galaxies, 
there is relationship between the host-gas temperature and its X-ray
luminosity (e.g. Fukazawa, Makishima \& Ohashi 2004 and references
therein). Using the relationship between these two quantities 
reported in Fukazawa, Makishima \& Ohashi (2004) for a compilation
of $\sim$300 objects, including elliptical galaxies, groups and clusters
of galaxies we have estimated the  
range of temperatures expected for a given X-ray luminosity and
compared it with the best-fit one. All the sources for which a
thermal model is acceptable have an X-ray luminosity below 
6$\times$10$^{43}$ erg s$^{-1}$ that corresponds to expected
temperatures below 3 keV. In all these cases, instead, the
best-fit temperatures range between 3 and 10 keV. These temperatures 
are usually observed only in very luminous clusters 
(L$_X$=10$^{44}$-10$^{46}$ erg s$^{-1}$). Even if we consider the
uncertainties on the kT values, the lower limits on the temperatures are 
too high considering the observed X-ray luminosities in all cases. 

In conclusion, a single thermal model, with physically acceptable 
temperatures, can be rejected confidently in all the objects.

%__________________________________________________________
%                                             Two column Table 
%_____________________________________________________________
%
\begin{table*}
\caption{X-ray spectroscopy}             
\label{tab_res_x}      
\begin{center}          
\begin{tabular}{l l r c r c c l r c}     % 7 columns 
\hline\hline       
name & z & net counts & model & $\Gamma$ & N$_H$ & kT & Red. $\chi^2$& Log L$_X$ & X-ray class\\
     &   &            &       &          &       &    &    /d.o.f.   &           &            \\     
     &   &    &  &        & [10$^{22}$ cm$^{-2}$] & [keV] & & [erg s$^{-1}$] & \\
\hline
XBSJ000532.7$+$200716  & 0.11900 &   1118 & BB+PL                &  2.32$^{+ 0.52}_{-0.30}$ & $<$   0.49 &   0.10 &  0.98 /   39 & 42.39 & nabs AGN \\
XBSJ012540.2$+$015752  & 0.12310 &    697 & PL                   &  1.79$^{+ 0.15}_{-0.12}$ & $<$   0.03 & ... &  1.35 /   37 & 42.62 & nabs AGN \\
XBSJ012654.3$+$191246  & 0.04268 &    154 & T+PL                 &  1.95$^{+ 0.71}_{-0.39}$ & $<$   1.20 &   0.99 &  1.14 /    5 & 41.25 & AGN \\
XBSJ013944.0$-$674909  & 0.10400 &    794 & PL                   &  1.94$^{+ 0.14}_{-0.12}$ & $<$   0.02 & ... &  0.91 /   34 & 42.56 & nabs AGN \\
XBSJ014109.9$-$675639  & 0.22600 &    408 & PL                   &  1.78$^{+ 0.39}_{-0.34}$ &    0.12$^{+   0.21}_{  -0.12}$ & (5.80) &  0.72 /   16 & 43.53 & nabs AGN \\
XBSJ014227.0$+$133453  & 0.27500 &    120 & PL                   &  1.80$^{+ 0.72}_{-0.53}$ &    0.15$^{+   0.38}_{  -0.15}$ & (8.80) &  0.70 /    3 & 43.44 & nabs AGN \\
XBSJ021822.2$-$050615$^1$ & 0.04400 &   1772 & Leaky+Line           &  1.66$^{+ 0.34}_{-0.36}$ &   20.54$^{+   0.36}_{  -0.44}$ & ... &  1.41 /   63 & 42.53 & abs AGN \\
XBSJ025645.4$+$000031  & 0.35850 &    399 & PL                   &  1.93$^{+ 0.37}_{-0.19}$ & $<$   0.14 & (4.10) &  1.15 /   15 & 43.48 & nabs AGN \\
XBSJ031859.2$-$441627$^1$ & 0.13954 &    350 & PL                   &  1.72$^{+ 0.43}_{-0.37}$ &    0.39$^{+   0.34}_{  -0.28}$ & ... &  0.80 /   14 & 42.99 & nabs AGN \\
XBSJ043448.3$-$775329  & 0.09700 &    244 & PL                   &  1.50$^{+ 0.12}_{-0.11}$ &    0.27$^{+   0.12}_{  -0.11}$ & ... &  1.23 /   33 & 42.86 & nabs AGN \\
XBSJ050453.4$-$284532  & 0.20400 &   1205 & PL                   &  1.45$^{+ 0.12}_{-0.14}$ &    0.04$^{+   0.04}_{  -0.04}$ & ... &  0.86 /   54 & 43.16 & nabs AGN \\
XBSJ051822.6$+$793208  & 0.05250 &    979 & PL                   &  1.86$^{+ 0.11}_{-0.11}$ & $<$   0.02 & ... &  0.90 /   43 & 41.83 & nabs AGN \\
XBSJ052116.2$-$252957  & 0.33200 &    151 & PL                   &  2.10$^{+ 0.75}_{-0.34}$ & $<$   0.32 & (3.30) &  0.74 /    6 & 43.22 & nabs AGN \\
XBSJ052128.9$-$253032  & 0.58750 &     60 & PL                   & 1.9 (frozen) &   12.71$^{+   6.46}_{  -3.98}$ & ... &  0.87 /    5 & 44.45 & abs AGN \\
XBSJ075117.9$+$180856$^1$ & 0.25500 &    913 & PL                   &  1.58$^{+ 0.16}_{-0.16}$ &    0.11$^{+   0.07}_{  -0.07}$ & ... &  0.86 /   42 & 43.50 & nabs AGN \\
XBSJ083737.0$+$255151  & 0.10520 &    261 & PL                   &  1.77$^{+ 0.35}_{-0.33}$ &    0.34$^{+   0.30}_{  -0.22}$ & ... &  0.75 /   14 & 43.00 & nabs AGN \\
XBSJ090729.1$+$620824  & 0.38800 &     90 & PL                   & 1.9 (frozen) &    0.90$^{+   1.58}_{  -0.62}$ & ... &  1.90 /    3 & 43.63 & abs AGN \\
XBSJ094526.2$-$085006  & 0.31400 &    136 & PL                   &  2.22$^{+ 1.24}_{-0.75}$ &    0.32$^{+   0.87}_{  -0.32}$ & (5.90) &  1.66 /    5 & 43.49 & nabs AGN \\
XBSJ100032.5$+$553626  & 0.21600 &    266 & Leaky                &  2.26$^{+ 0.42}_{-0.29}$ &   25.70$^{+  84.60}_{ -21.10}$ & ... &  0.86 /   13 & 43.93 & abs AGN \\
XBSJ101843.0$+$413515  & 0.08400 &    837 & PL                   &  1.87$^{+ 0.18}_{-0.11}$ & $<$   0.04 & ... &  1.11 /   41 & 42.30 & nabs AGN \\
XBSJ111654.8$+$180304  & 0.00312 &    822 & T+PL                 &  2.25$^{+ 0.37}_{-0.25}$ & $<$   0.07 &   0.71 &  0.66 /   34 & 39.00 & GAL \\
XBSJ112026.7$+$431520  & 0.14600 &    412 & Leaky+Line           &  1.65$^{+ 0.83}_{-0.42}$ &    6.28$^{+   3.21}_{  -1.73}$ & ... &  0.88 /   21 & 43.20 & abs AGN \\
XBSJ122017.5$+$752217  & 0.00586 &   8190 & T+PL                 &  2.35$^{+ 0.55}_{-0.45}$ &    1.45$^{+   0.81}_{  -0.63}$ &   0.51 &  0.99 /  231 & 40.20 & GAL? \\
XBSJ133942.6$-$315004  & 0.11409 &   1076 & PL                   &  1.62$^{+ 0.22}_{-0.18}$ &    0.23$^{+   0.12}_{  -0.08}$ & ... &  1.32 /   50 & 42.81 & nabs AGN \\
XBSJ134656.7$+$580315  & 0.37300 &    106 & PL                   & 1.9 (frozen) &    9.50$^{+   4.72}_{  -3.16}$ & ... &  1.92 /   13 & 44.04 & abs AGN \\
XBSJ140100.0$-$110942  & 0.16400 &   1305 & PL                   &  2.33$^{+ 0.11}_{-0.10}$ & $<$   0.02 & ... &  1.23 /   57 & 42.57 & nabs AGN \\
XBSJ142741.8$+$423335  & 0.14200 &    487 & PL                   & 1.9 (frozen) &    4.48$^{+   0.93}_{  -0.77}$ & ... &  0.85 /   23 & 43.20 & abs AGN \\
XBSJ143835.1$+$642928  & 0.11800 &    404 & PL                   &  1.84$^{+ 0.42}_{-0.18}$ &    1.85$^{+   0.73}_{  -0.55}$ & ... &  0.73 /   23 & 43.03 & abs AGN \\
XBSJ143911.2$+$640526  & 0.11290 &    123 & PL                   & 1.9 (frozen) &   20.00$^{+   9.72}_{  -6.56}$ & ... &  0.52 /    6 & 42.96 & abs AGN \\
XBSJ161820.7$+$124116  & 0.36100 &     40 & PL                   & 1.9 (frozen) &    4.96$^{+   6.14}_{  -2.48}$ & ... &  1.17 /    3 & 43.72 & abs AGN \\
XBSJ163332.3$+$570520  & 0.38600 &    613 & PL                   &  2.19$^{+ 0.72}_{-0.39}$ &    0.03$^{+   0.29}_{  -0.03}$ & (3.70) &  0.78 /   16 & 43.45 & nabs AGN \\
XBSJ193248.8$-$723355  & 0.28700 &    782 & PL                   &  1.48$^{+ 0.25}_{-0.22}$ &    0.73$^{+   0.35}_{  -0.29}$ & ... &  0.86 /   33 & 43.80 & abs AGN \\
XBSJ230401.0$+$031519  & 0.03630 &    183 & PL                   &  1.78$^{+ 0.88}_{-0.24}$ & $<$   0.15 & (4.30) &  0.95 /    5 & 41.47 & nabs AGN \\
XBSJ230434.1$+$122728  & 0.23200 &    180 & PL                   &  1.52$^{+ 0.43}_{-0.37}$ & $<$   0.18 & ... &  1.08 /    5 & 43.31 & nabs AGN \\
XBSJ231546.5$-$590313$^2$ & 0.04460 &    833 & T+Leaky              &  1.67$^{+ 0.09}_{-0.34}$ &    6.90$^{+   4.70}_{  -3.40}$ &   0.65 &  0.74 /   44 & 42.04 & abs AGN \\

\hline
\end{tabular}
\end{center}
column 1: name ($^1$ = X-ray analysis taken from Severgnini et al. 2003; $^2$ = X-ray analysis taken from Franceschini et al. 2003) 

column 2: redshift

column 3: net counts used for the spectral analysis

column 4: best-fit model (PL=single absorbed power-law model; BB=black-body emission; T=Thermal (Mekal) model;
Leaky = Leaky model (see text for details)

column 5: power-law photon index

column 6: intrinsic absorption column density

column 7: best-fit temperature of the additional thermal component or, if in parenthesis,
the temperature of the thermal model (alternative to the power-law model, see text for details)

column 8: reduced $\chi^2$ and degrees of freedom of the best-fit model

column 9: Log of the de-absorbed 2-10 keV X-ray luminosity

column 10: classification based on the X-ray analysis: nabs AGN = non-absorbed (or 
partly absorbed) AGN (N$_H<$4$\times$10$^{21}$ cm$^{-2}$); abs AGN = absorbed
AGN  (N$_H>$4$\times$10$^{21}$ cm$^{-2}$); GAL =  X-ray from normal galaxy (thermal and/or
from X-ray binaries) 
\end{table*}

In Tab.~\ref{tab_res_x} we summarize the results of the X-ray analysis. 
The best-fit model is indicated in column~3. A value of temperature 
(column~6) reported between parenthesis means that a thermal model
is, from a statistical point of view,  alternative to the one indicated
in column~3 but has been rejected as ``unphysical'' on the basis
of the considerations discussed above.

\subsection{X-ray evidence for AGN?} 
The presence of a power-law (or two power-laws) emission model in all
the sources discussed here can be considered as suggestive for
the presence of an AGN in all cases. It should be 
noted, however, that in the very low-luminosity regime 
(L$_X<$ a few 10$^{40}$) hard X-ray emission is often detected in
elliptical galaxies, besides the usual thermal emission associated
with the hot gas (kT$\sim$1 keV) (e.g. Matsushita et al. 1994). The origin
of this emission has been attributed mainly to discrete binary X-ray sources 
(Matsushita et al. 1994). The confirmation that many ellipticals actually
harbor a population of point-like sources with luminosities ranging 
from 10$^{37}$ to 10$^{39}$ erg s$^{-1}$ or even (but rarely) above 
10$^{39}$ erg s$^{-1}$, has been produced by high resolution 
{\it Chandra} images (see Fabbiano \& White 2006, for a review). 
The integrated X-ray luminosity of the X-ray binaries in elliptical and
S0 galaxies typically ranges from 6$\times$10$^{39}$ to 
9$\times$10$^{40}$ erg s$^{-1}$ (Kim \& Fabbiano 2004).
For this reason, caution must be paid before considering the 
presence of a power-law fit as suggestive of the presence of an AGN
in the low-luminosity regime ($<$10$^{41}$ erg s$^{-1}$).
Only two objects in our sample, 
XBSJ111654.8+180304 (NGC3607) and XBSJ122017.5$+$752217 (NGC4291) 
have 2-10 keV luminosities below 10$^{41}$ erg s$^{-1}$. 

XBSJ111654.8+180304 (NGC3607) is a LINER (Ho et al. 1997). 
The X-ray luminosity 
(1.1$\times$10$^{39}$ erg s$^{-1}$) is the lowest among the 
extragalactic XBS sources. The X-ray spectrum is well fitted
by a steep power-law model ($\Gamma$=2.3) plus a thermal component 
with kT=0.71 keV. Based on Chandra (and HST) data, Gonz\'alez-Mart\'in
et al. (2006) have concluded that the observed hard X-ray emission is
extended and not related to a nuclear source.

XBSJ122017.5$+$752217 (NGC4291) is a
galaxy with an early-type spectrum with no emission lines. Its 
X-ray luminosity (2$\times$10$^{40}$ erg s$^{-1}$) is 
similar to those observed by Matsushita et al. (1994) and the X-ray spectral 
properties (a soft thermal component with kT=0.51 keV plus a hard tail) 
could be either indicative of the presence of a very low-luminosity AGN or
of the presence of a population of X-ray binaries. 
Observations in the soft X-ray with high spatial resolution, 
carried out by  ROSAT High Resolution Imager (HRI), seem to favour the 
low-luminosity AGN hypothesis (Roberts \& Warwick 2000; Liu \& Bregman 2005)
but the conclusion is uncertain. We tentatively classify this object as 
``non-AGN'' and we do not consider it as AGN in the following
analysis.

All the remaining sources have X-ray luminosities larger than 10$^{41}$ erg 
s$^{-1}$ and  are well fitted only by a power-law 
model or a leaky model (in a few cases plus an additional thermal  
emission) and we consider these elements as strongly
indicative of the presence of an AGN.

It must be noted that some of the 35 objects analysed here have an optical 
spectrum characterized by narrow emission lines suggestive of star-formation.
Higher X-ray luminosities are expected in these cases when compared to
``passive'' elliptical galaxies. The most extreme cases are represented by
the Ultraluminous Infrared Galaxies (ULIRG) where 2-10 keV 
luminosities\footnote{converted into the cosmology adopted here}  
up to 2-3$\times$10$^{41}$ erg s$^{-1}$
are 
often observed even if an AGN
is not clearly present (e.g. Franceschini et al. 2003). All the objects in our  sample
with a starburst-like  spectrum have 2-10 keV luminosities above 10$^{42}$ erg s$^{-1}$ 
and it is unlikely that it is all due to the starburst activity. For instance, the 
most extreme starburst in the sample (the ULIRG  XBSJ231546.5-590313) has been
studied in detail by Franceschini et al. (2003) and the AGN component has been
clearly detected out of the starburst emission in the X-ray spectrum.

In conclusion, the X-ray spectral analysis has revealed in all but two sources
the presence of an AGN which is either absorbed or unabsorbed/mildly absorbed. 
This result confirms 
what has been already estimated from the X-ray-to-optical 
flux ratios and the observed X-ray luminosities i.e. that most 
of the 35 objects discussed here actually contain an AGN.

Given the impossibility of using the optical spectra for a reliable classification of
the elusive AGN we have decided to use the results of the X-ray analysis to classify these
objects. In order to match the optical classification into type~1 and type~2 AGN adopted 
for the XBS sources, that corresponds to a dividing limit on the optical absorption of 
A$_V$ of $\sim$2 magnitudes, as 
discussed in Caccianiga et al. (2007, in prep.), we have used the corresponding value of 
N$_H$=4$\times$10$^{21}$ cm$^{-2}$ (assuming the Galactic standard N$_H$/A$_V$ relation) 
to separate the elusive AGNs into type~1 and type~2.
With this threshold we thus classify 11 objects as ``type~2'' AGN and 20 objects as
type~1 AGN. We have considered as non-absorbed also an object (XBSJ000532.7$+$200716) for which
we have determined an upper limit on N$_H$ which is marginally higher than 4$\times$10$^{21}$ cm$^{-2}$.
Finally, for one source (XBSJ012654.3$+$191246)  the upper limit on N$_H$ is not stringent enough
($<$1.2$\times$10$^{22}$ cm$^{-2}$) to allow a classification of the source.

\section{Why are they hidden in the optical?}
The X-ray analysis discussed in the previous section has demonstrated
the presence of an AGN in all but two sources studied here. We now want
to investigate the reason (or the reasons) why the AGN is hardly
detected (or not detected at all) in the optical band. 

Different hypothesis have been discussed in the recent literature to explain the
presence of a large number of ``optically dull'' sources in X-ray surveys.
Comastri et al. (2002, 2003), for instance, have suggested that an heavily
absorbed AGN (Compton thick with N$_H>$10$^{24}$ cm$^{-2}$) could be hidden within
some XBONG. This hypothesis is suggested by the fact that, in the magnitude/X-ray flux
plot, the XBONG discovered in {\it Hellas2XMM} survey occupy a region where also
Compton thick AGN are expected to be found, given their typical spectral energy
distribution and redshift (Comastri et al. 2003). 
Alternatively, it has
been proposed by different authors (Moran et al. 2002;
Severgnini et al. 2003; Georgantopoulos \& Georgakakis 2005; Page et al. 2006) that the elusiveness
of many X-ray selected AGN is just the consequence of the intrinsic  weakness 
(in respect to the host-galaxy) combined in some cases with moderate absorption 
(10$^{22}$ cm$^{-2}$-10$^{24}$ cm$^{-2}$) of an otherwise ``normal'' AGN. This effect, 
often called ``optical dilution'', is expected when the optical spectra are taken
with a relatively large slit width including a large amount of star-light
from the host-galaxy.
Another possible explanation is that the 
``optically-dull'' AGN (or at least a fraction
of them) are characterized by unusual properties like a weak emission from the accretion 
disk (e.g. a radiatively inefficient accretion flow RIAF, Yuan \& Narayan (2004)
that produces  extremely ``flat''  values of  
$\alpha_{OX}$\footnote{The two point spectral index $\alpha_{OX}$ is defined
as: $\alpha_{OX} = -\frac{Log(f_o/f_x)}{Log(\nu_o/\nu_x)}$ where $f_o$ and 
$f_x$ are, respectively, the monochromatic fluxes at 
$\nu_o$=1.20$\times$10$^{15}$ Hz (corresponding to $\lambda_o$=2500\AA) and
$\nu_x$=4.84$\times$10$^{17}$ Hz (corresponding to E=2 keV).
} 
spectral index. Similarly, a BL Lac object can be a valid possibility (Brusa et al. 2003)
although, given the low space density of these objects, it is unlikely that this
explanation can account for all (or a large fraction) of the ``elusive'' AGN selected in
a X-ray survey.
Finally, the entire AGN (including the Narrow Line Region) could suffer 
from extra-torus absorption, located, for instance, on large scales within the host 
galaxy (e.g. Rigby et al. 2006; Cocchia et al. 2007). Rigby et al (2006), 
in particular, consider that, at high-redshifts ($>$1) besides the host-galaxy 
dilution the extra-nuclear dust in the host galaxy is likely to contribute to 
hide the AGN.

In the following sections we analyse all these competing hypotheses to assess their 
relative importance to explain the ``elusive AGN'' found in the XBS survey.

\subsection{Compton-thick hypothesis}
The detection of a Compton-thick AGN, using the 2-10 keV energy band may be missed,
in particular for local AGN, since the energy cut-off is expected to fall outside the
observed interval. Other indicators can be used to assess its presence, like the
detection  of a prominent Fe K$\alpha$ emission line at 6.4 keV or a very low ($<$1)
F$_X$/F$_{[OIII]}$ flux ratio. This latter indicator, called Compton-thickness parameter (T),
is relatively easy to apply and its reliability has been already tested (Bassani et al. 1999)
at least for local AGN. The indicator is based to the fact that, in presence of a high
level of absorption, the 2-10 keV flux is expected to be heavily suppressed while the
[OIII]$\lambda$5007\AA\ should be unaffected. The available X-ray spectra of the elusive
XBS AGNs do not have - on average - enough counts to asses the presence of the 
K$\alpha$ line or to compute a stringent upper limit. For this reason we use the
T parameter to test the  Compton-thick hypothesis.

The optical spectra of the XBS sources have been 
taken with the primary aim of deriving the spectral classification and the 
redshift.
Since both quantities do not require a reliable absolute calibration, the 
observing runs
are not necessarily carried out under photometric conditions 
and no corrections have been applied, for instance,
for the flux-loss due to the width of the slit. Using the magnitudes available 
for the XBS objects we have statistically corrected the fluxes in each 
observing run and estimated the uncertainty on the final flux which turned 
out to have a relative error of $\sim$50\% (see Appendix A for details).

In Figure~\ref{thickness} we report
the Compton-thickness parameter as a function of the X-ray flux for all the 35 
objects discussed in this paper i.e. including the 2 sources classified as normal galaxies. 
We have corrected the [OIII]$\lambda$5007\AA\ fluxes for the
extinction due to our Galaxy.
The extinction due to the host galaxy is more difficult to 
estimate given the type of spectra under analysis (e.g. the Balmer decrement 
cannot be measured). However we can evaluate the maximum level of extinction 
that can be expected given the type of galaxies under analysis. The largest levels of
extinction are expected for galaxies with high star-forming where extinctions up
to A$_V\sim$1 can be observed (Calzetti \& Heckman 1999). In the majority of the
elusive AGN discussed here, however, the host-galaxy does not show evidence of high star-formation rate
given the absence (or the weakness) of emission lines and, therefore, the extinction is
expected to be well below 1 magnitude. Only for the few emission lines galaxies in the
sample the extinction could reach A$_V\sim$1 (the most extreme case is XBSJ231546.5-590313, an 
ULIRG whose optical spectrum is dominated by the star-formation). The maximum variation of
the Compton-thickness parameter expected in these cases is reported in the box located
in the upper-left side of  Figure~\ref{thickness}.

From Figure~\ref{thickness} it is clear that all  the elusive sources in the XBS survey 
have a Compton-thickness parameter well above 1. Considering the maximum level of extinction 
($\sim$1 mag) due to the host-galaxy only the three objects with T$<$10 fall close (but not within) 
the Compton-thick region. Two of these sources are tentatively classified as ``normal'' galaxies 
(XBSJ111654.8+180304 and XBSJ122017.5$+$752217, see discussion in Section~3.1) although 
in one of these (XBSJ111654.8+180304, NGC3607) 
the actual absence/presence of an AGN is still debated. The third source (XBSJ012654.3+191246)
has been classified as AGN on the basis of the X-ray emission but its further characterization as 
absorbed or unabsorbed is not possible given the high upper limit on the intrinsic value
of N$_H$. The 2-10 keV luminosities of these three sources are the lowest observed in the XBS
survey (between 10$^{39}$ and 2$\times$10$^{41}$ erg s$^{-1}$). 
At present we can rule out the Compton-thick hypothesis for XBSJ111654.8+180304 for which
high-resolution X-ray data (Chandra) have not detected any hard point-like nucleus in this
object (Gonz\'alez-Mart\'in et al. 2006). For the remaining two sources we need additional
data (e.g. high S/N X-ray spectrum to detect or rule out the presence of the 
Fe K$\alpha$ emission line). 

We conclude that the Compton-thick
hypothesis is not supported for most (if not all) of the elusive AGN in the XBS survey. 
Indeed, the presence of a Compton thick
AGN  should produce strong narrow emission lines in the optical spectrum
of sources with bright X-ray fluxes ($>$10$^{-13}$ erg s$^{-1}$ cm$^{-2}$) something that, 
by definition, is not found among the elusive objects.
%                                     One column figure (place early!)
%______________________________________________ Gamma_1 (lg rho, lg e)
   \begin{figure}
   \centering
    \includegraphics[width=9cm]{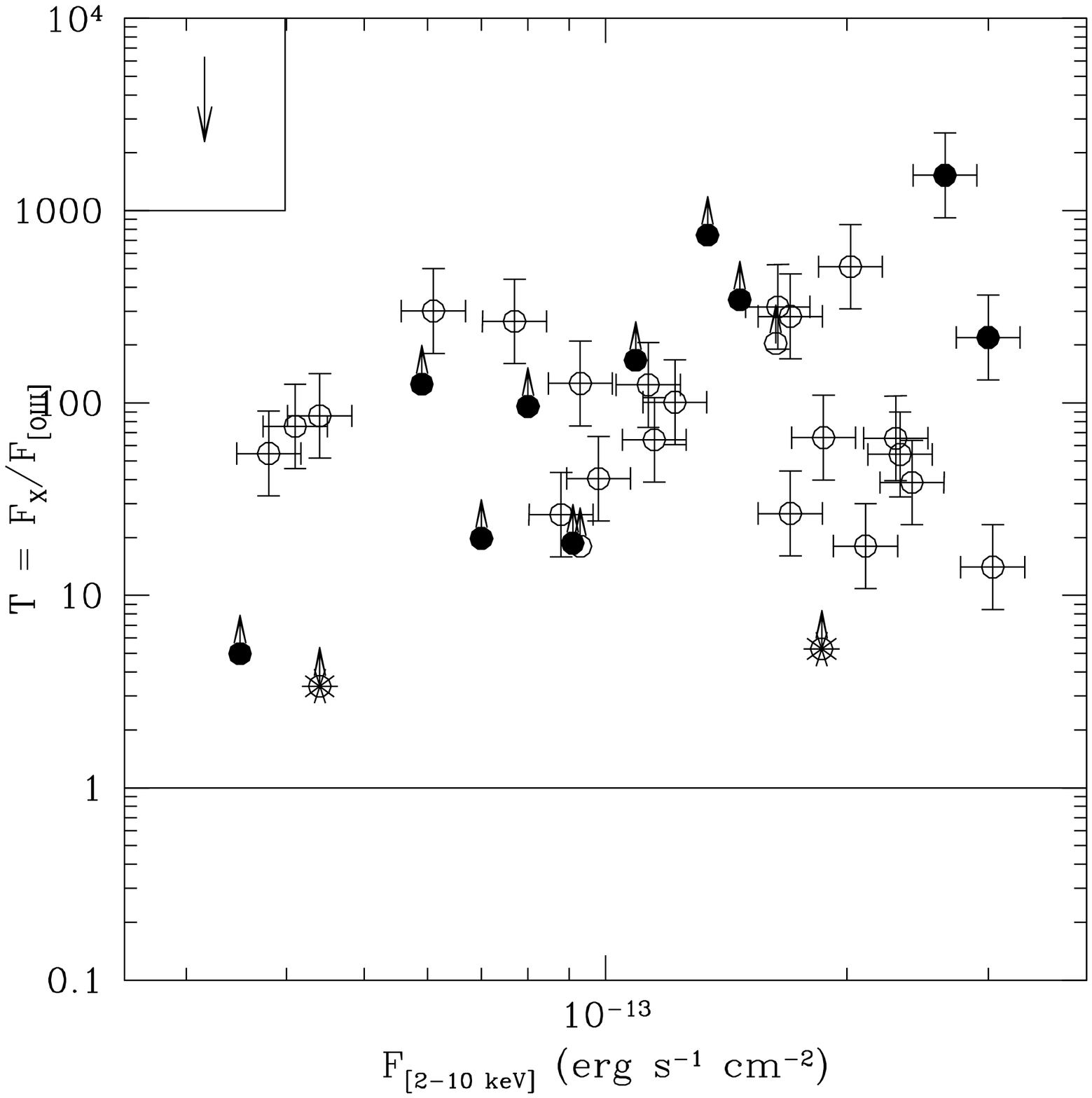}

   \caption{The Compton-thickness parameter (corrected for Galaxy extinction) 
versus the X-ray flux of the elusive AGN and normal galaxies of the XBS sources. 
Circles represent the sources classified as AGN from the X-ray analysis. Filled circles, in
particular, are the sources with an early-type spectrum (XBONG). The two stars 
are the sources that probably do not contain an AGN (see text for details). 
The arrow within the box in the upper-left side of the figure represents the maximum 
variation of T due to the host-galaxy extinction.
}
              \label{thickness}
    \end{figure}

\subsection{Weakness/moderate absorption}

\begin{figure}
   \centering
    \includegraphics[width=9cm]{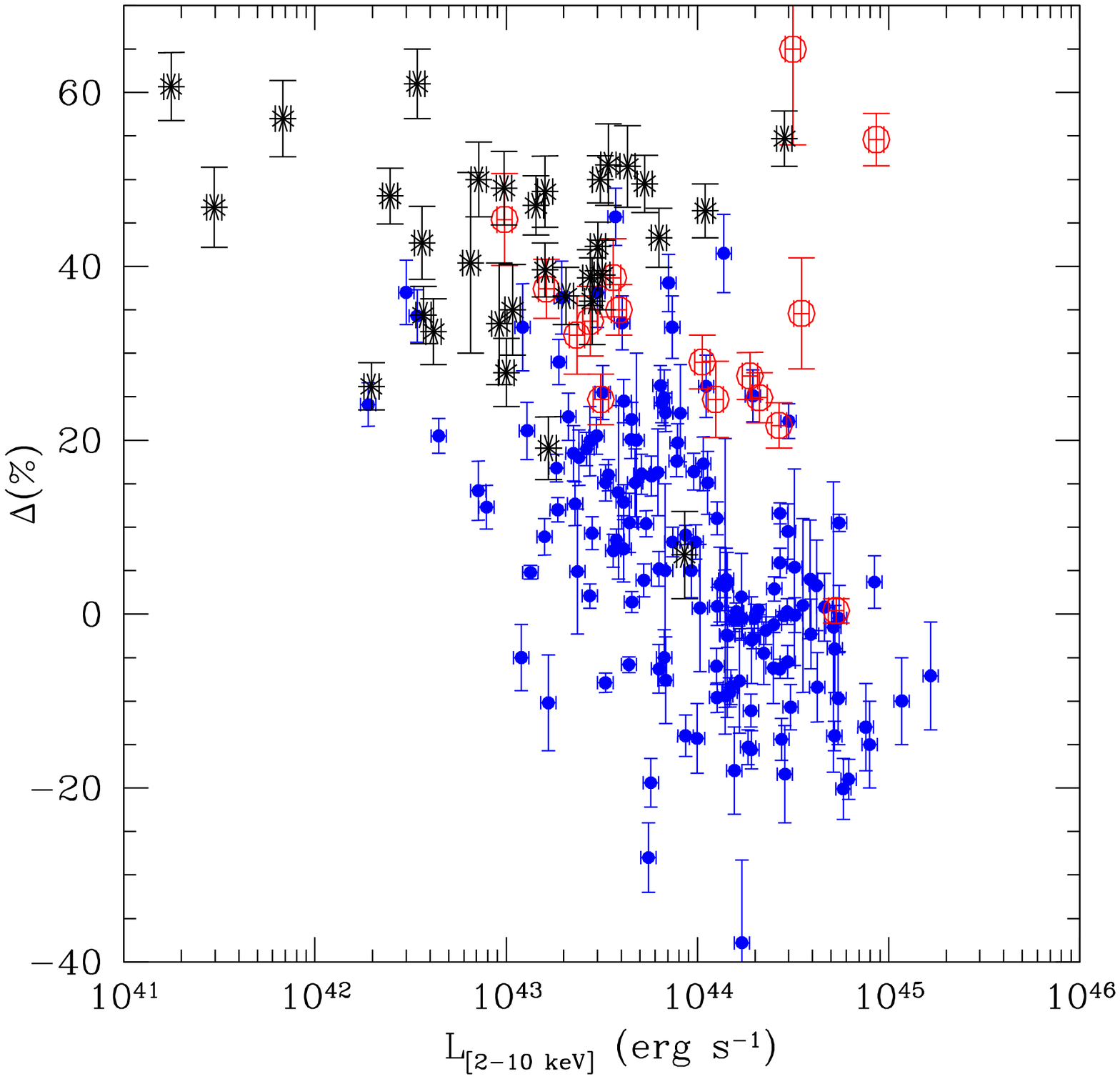}

   \caption{The 4000\AA\ break ($\Delta$), given in
percentage, versus the  de-absorbed 2-10 keV 
X-ray luminosity for the XBS sources. 
Filled points are type~1 AGN, open circles are
type~2 AGN, stars are elusive AGN.
}
              \label{lx_c}%
    \end{figure}

In Fig.~\ref{lx_c} we report the 4000~\AA\ break
($\Delta$\footnote{The 4000\AA\ break is
defined as $\Delta$ = $\frac{F^+ - F^-}{F^+}$ where F$^+$ and F$^-$ 
represent the mean value of the flux density 
(expressed per unit  frequency) in the region 4050 - 4250~\AA\ and
3750 - 3950~\AA\ (in the source's rest-frame) respectively.})
versus 
the unabsorbed X-ray luminosities of all the XBS objects. 
The 4000~\AA\ break is
a rough indicator of the importance of the galaxy star-light
in the total emission of the source. When the nuclear emission
is dominant (at $\sim$4000\AA, source's rest-frame), the
value of $\Delta$ is just an indicator of the slope of the nuclear
emission: the bluer the spectrum, the lower (typically negative)
the value of $\Delta$. In these cases the value of $\Delta$ is usually 
below 20\%. When, on the contrary, the continuum is dominated 
by the host galaxy light the value of $\Delta$ ranges from 20\% up to 60\%, 
depending on the type of host-galaxy. In the 20\%-40\% range of $\Delta$ we
find also sources where both the nucleus and the host galaxy contribute
to the observed spectrum.

The X-ray luminosity is an independent indicator of the AGN strength,
at least in the 33 sources for which the presence of an AGN has been
assessed (the 2 ``normal'' galaxies are not plotted in this figure).
%                                     One column figure (place early!)
%______________________________________________ Gamma_1 (lg rho, lg e)
%                                     One column figure (place early!)
%______________________________________________ Gamma_1 (lg rho, lg e)
   \begin{figure}
   \centering
    \includegraphics[width=9cm]{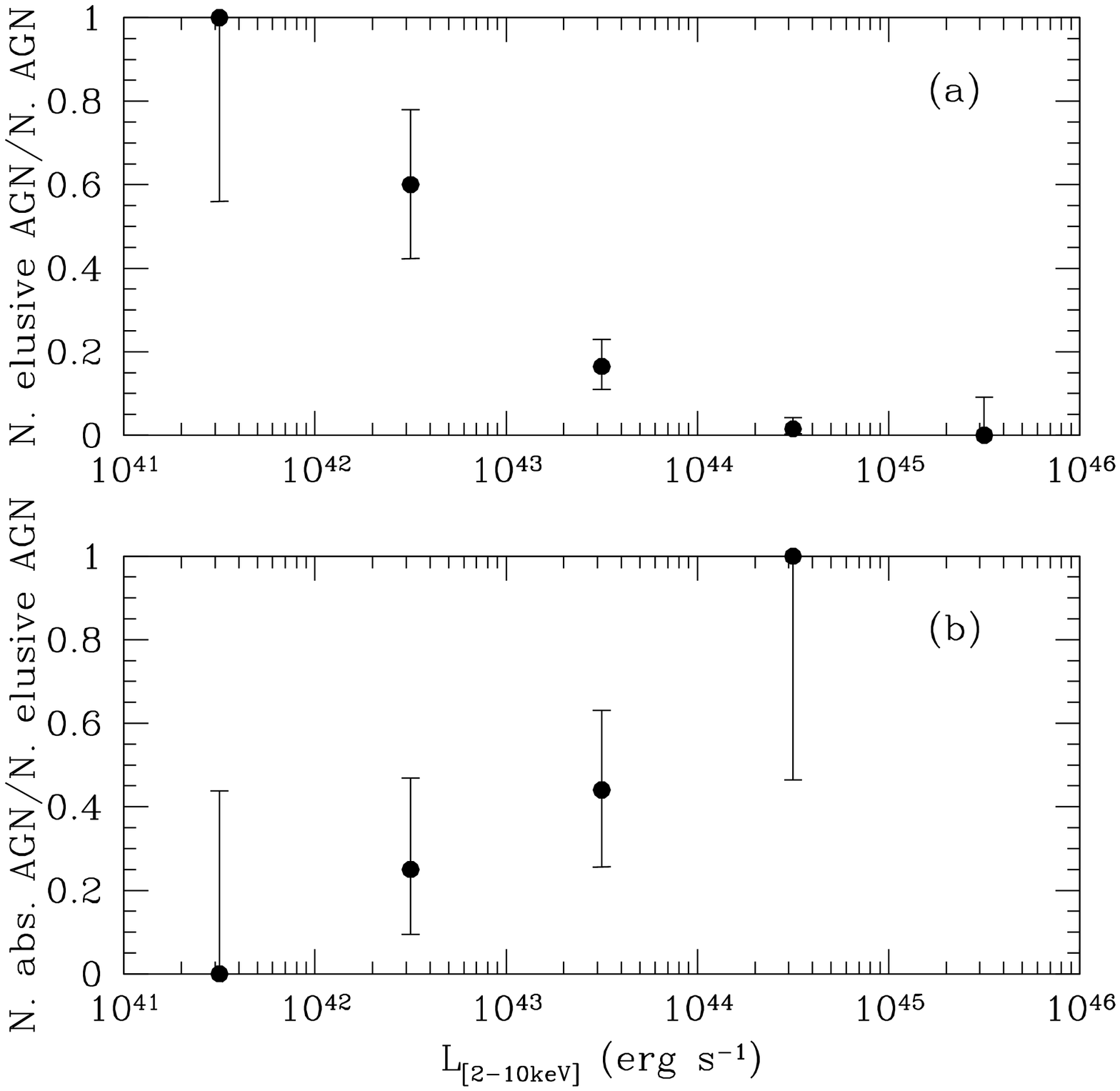}

   \caption{(a) Fraction of elusive AGN as a function of the
2-10 keV de-absorbed X-ray luminosity; (b) Fraction of absorbed AGN
among the elusive AGN. The error-bars are computed using
a Bayesian method: in particular, we plot the shortest
90\% confidence interval.
              \label{frac}}
    \end{figure}

 Fig.~\ref{lx_c} shows an evident anti-correlation between $\Delta$ and the 
X-ray luminosity: powerful (L$_X>$10$^{44}$ erg s$^{-1}$) AGN are only
marginally affected by the recognition problem,
and the host-galaxy contribution
to the total emission at 4000~\AA\ is usually low or negligible 
($\Delta<$30\%); 
for X-ray luminosities between 10$^{43}$ and 10$^{44}$ 
erg s$^{-1}$ the fraction of elusive AGNs becomes more important
and the values of $\Delta$ range from -30 to 50\%; in the low-luminosity
regime ($<$10$^{43}$ erg s$^{-1}$) the majority of the sources is
optically elusive and $\Delta$ is in the typical
range of values observed in galaxies (20\%-60\%). 
This trend is shown in Figure~\ref{frac}a which 
reports the fraction of elusive AGN as a function
of the de-absorbed 2-10 keV X-ray luminosity. 

The observed trend is a first indication supporting the  hypothesis 
that optically elusive AGN are simply the low-luminosity tail of the 
AGN population. 

In Fig.~\ref{frac}b the fraction of elusive AGN classified as ``absorbed'' AGN
(N$_H>$4$\times$10$^{21}$ cm$^{-2}$) according to the X-ray analysis has been
reported. For X-ray luminosities below 10$^{43}$ erg s$^{-1}$ the fraction
of absorbed AGN among the elusive AGN is similar to the fraction of 
absorbed AGN among the total sample of AGN in the XBS survey ($\sim$10\%-30\% 
in the BSS and HBSS samples respectively). 
For higher X-ray 
luminosities, instead,  the fraction of absorbed objects among elusive AGN 
increases significantly 
and becomes dominant for luminosities $>$10$^{44}$ erg s$^{-1}$. The obvious
interpretation of this trend is that, beside the luminosity of the AGN, 
also the absorption (but still in Compton-thin regime) plays an important role to 
``hide'' the 
AGN and, for high X-ray luminosities ($>$10$^{44}$ erg s$^{-1}$), this becomes the 
most important explanation.

%                                     One column figure (place early!)
%______________________________________________ Gamma_1 (lg rho, lg e)
   \begin{figure}
   \centering
    \includegraphics[width=9cm]{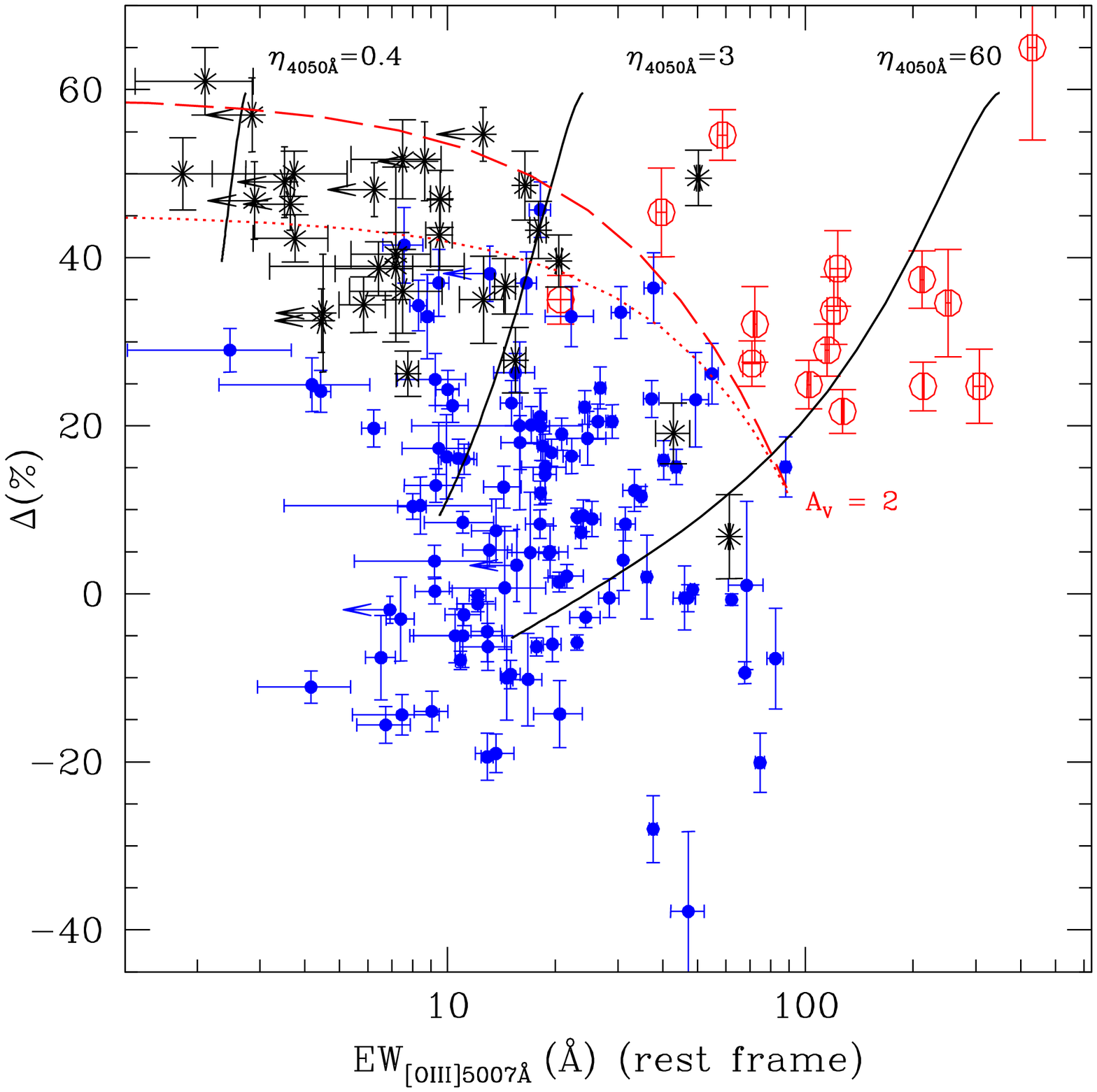}

   \caption{The 4000\AA\ break ($\Delta$), given in
percentage, versus the [OIII]$\lambda$5007\AA\ equivalent
width of the type~1 AGN (filled points), type~2 AGN (open circles) 
and the optically  elusive AGN (stars). The 3 solid lines 
labeled as $\eta_{4050\AA}$ = 0.4, 3 and 60 
show the expected ``paths'' followed when the optical absorption 
increases from 
0 to A$_V\sim$500 in sources with an intrinsic AGN/host-galaxy flux
ratio ($\eta_{4050\AA}$) of 0.4, 3 and 60 respectively. The dashed
line shows the region corresponding to an absorption
level of A$_V$=2 mag. The dotted line shows how the A$_V$=2 mag 
line changes when a younger host galaxy is assumed (t=1 Gyr instead
of 10 Gyr).
}
              \label{loiii_c}%
    \end{figure}

In Fig.~\ref{loiii_c} we show the values of 4000~\AA\ break versus the 
[OIII]$\lambda$5007\AA\ equivalent widths for the type~1, type~2 and the 
optically elusive AGN in the XBS survey. The [OIII]$\lambda$5007\AA\ 
emission line is a good indicator of the AGN activity, being unaffected
by the absorption (according to the simplest version of the unified models). 
On the same figure we 
plot the expected ``paths'' (continuous lines) 
of a source that is absorbed from 0 
up to A$_V\sim$500 (corresponding to N$_H\sim$10$^{24}$ assuming
the Galactic standard A$_V$/N$_H$ conversion). 
To produce these curves we have used the spectral model discussed in 
Severgnini et al. (2003).  This model  
includes a galaxy and an AGN template. 
The AGN template, in particular, is composed by two parts: a) 
the continuum with the broad emission lines and b) the narrow emission lines.
According to the basic version of the AGN unified model, the first part can 
be absorbed while the second one is not affected by the presence of
an obscuring medium. 
The AGN template is based on the data taken 
from Francis et al. (1991) and Elvis et  al. (1994) while the extinction curve 
is taken from Cardelli, Clayton \& Mathis (1989).
The galaxy template is produced on the basis of the Bruzual \& Charlot (2003)
models. 

In Fig.~\ref{loiii_c} we have considered
3 examples of curves where the intrinsic (i.e. not absorbed)
AGN/galaxy flux ratio at 4050\AA\ ($\eta_{4050\AA}$\footnote{
We define $\eta_{4050\AA}=\frac{f_{AGN}(4050\AA)}{f_{Gal}(4050\AA)}$
where $f_{AGN}$(4050\AA) and $f_{G}$(4050\AA) are the monochromatic
fluxes at 4050\AA\ (source's rest-frame) that pass through the slit of
the spectrograph.
}) is, respectively,
$\sim$60, 3 and 0.4. On the same plot we show the region corresponding
to an absorption of A$_V$=2 mag (dashed line). 
To produce these curves we have assumed
an host galaxy of t=10~Gyr. To show the impact of this assumption on the
plotted curves we report on the same figure also the A$_V$=2 mag
line computed assuming a much younger host galaxy (t=1 Gyr instead
of 10 Gyr, dotted line). 

Figure~\ref{loiii_c} confirms and quantifies what has been already suggested
by Figure~\ref{lx_c}: when the AGN is intrinsically powerful, if compared
to the host galaxy (e.g. $\eta_{4050\AA}>$3), optically absorbed and 
unabsorbed AGNs can be clearly detected and 
separated on the basis of the usual spectral 
classification criteria (see Caccianiga et al. 2007 in prep.) in the large 
majority of cases ($\sim$95\%).  When the AGN is 
optically weak compared to the host galaxy ($\eta_{4050\AA}<$3) the 
optical classification becomes 
difficult or impossible in $\sim$55\% of the cases. This percentage reaches 
$\sim$100\% when the  absorption level is high (A$_V>$2). 

The discussion presented above supports the idea that the intrinsic weakness of the 
AGN is the major reason for the optical dullness of the observed spectrum, at least
for X-ray luminosities below 10$^{43}$ erg s$^{-1}$. 
The presence of absorption (N$_H<$10$^{22}$ cm$^{-2}$) 
is an additional element that further reduces the
probability of detecting the AGN in the optical spectrum making the problem
of dilution more important for type~2 AGN ($\sim$40\% are elusive) in respect to
type~1 AGN (8\% are elusive). For X-ray luminosities higher than 10$^{44}$ erg s$^{-1}$, instead, 
the absorption is the most important reason for the AGN elusiveness.
The result that ``optical dilution'', besides the absorption, is an important 
ingredient for the lack of broad emission lines in many low-luminosity (L$_X<$10$^{43}$ erg s$^{-1}$)
AGNs has been recently suggested by Page et al. (2006), thus indicating that this
result holds also at fainter (10$^{15}$-5$\times$10$^{-14}$ erg s$^{-1}$ cm$^{-2}$)  X-ray fluxes.

\subsection{Intrinsically flat $\alpha_{OX}$ or host-galaxy absorption hypothesis}
We want to test here whether the XBS elusive sources (or a fraction  of them)
show a particularly ``flat'' value of  $\alpha_{OX}$ ($<$1) or unusual luminosities
of the Narrow-Line Region (when compared to the X-ray luminosity).
To this end, we first compute the luminosity of the [OIII]$\lambda$5007\AA\ 
emission line which is generally considered as a good tracer of the intrinsic 
AGN power. 

\begin{figure}
   \centering
    \includegraphics[width=9cm]{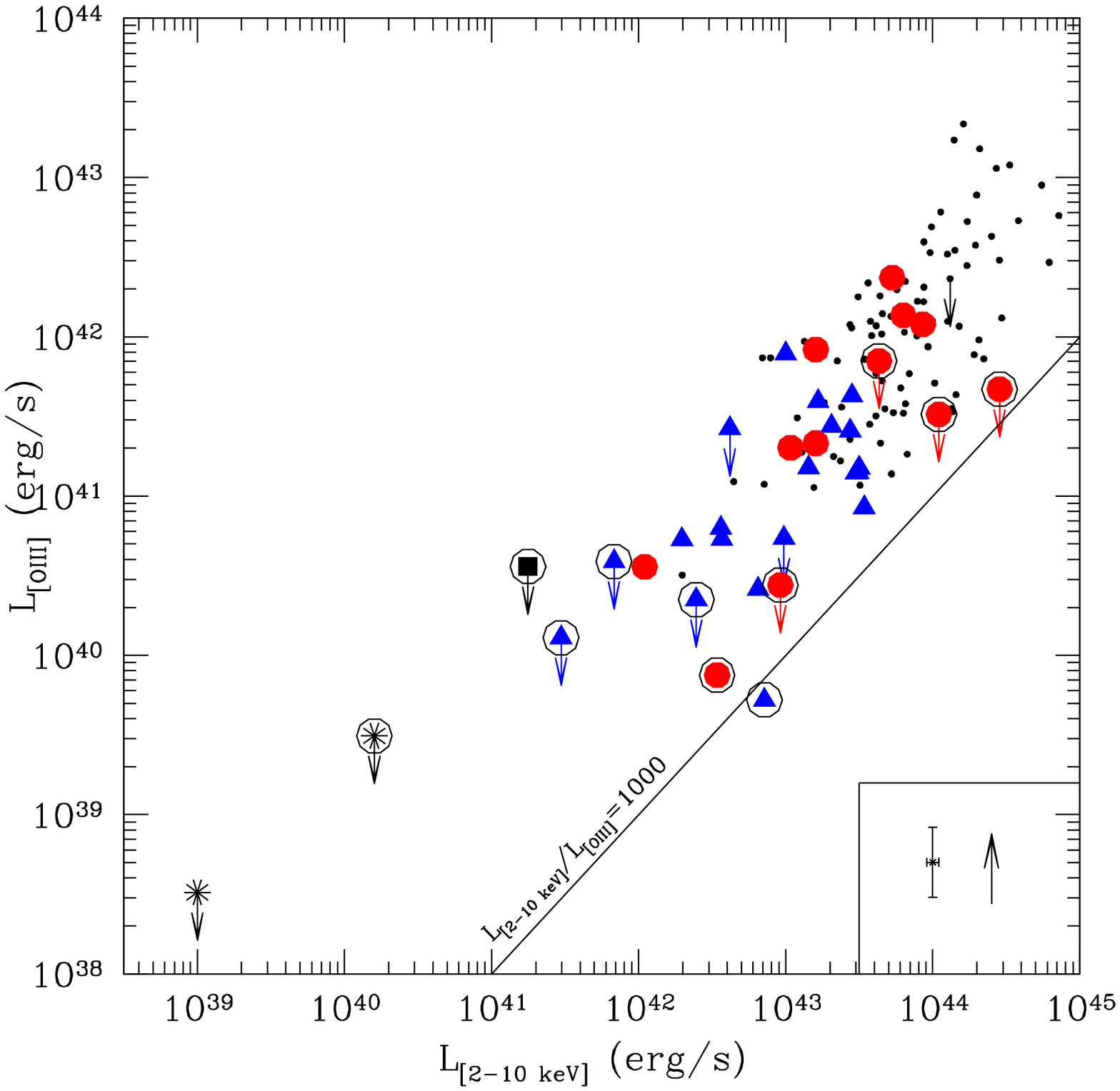}

   \caption{The [OIII]$\lambda$5007\AA\ luminosity versus the 
de-absorbed 2-10 keV X-ray luminosity for the type~1 AGN of the XBS
survey (small points), the elusive AGN (filled points=type 2 AGN,
filled triangles=type 1 AGN, filled box=AGN without further characterization into
type~1 or type~2) and the 2 normal galaxies (stars). The arrows 
correspond 
to the sources where the [OIII] emission line is not detected. 
The sources with an early-type spectrum (i.e. XBONG-like) are circled. 
The continuous line 
is the limit proposed by Cocchia et al. (2007) to classify an object
as XBONG. For clarity, we have reported
the typical error-bar in the right-bottom box. The arrow within this box
indicates the expected maximum correction for extinction due to the host-galaxy
(see discussion in Section~4.1 for details).
}
              \label{lumlum}
    \end{figure}

In Figure~\ref{lumlum} the [OIII]$\lambda$5007\AA\ luminosity
is plotted against the de-absorbed X-ray luminosity. We plot both the
non-elusive type~1 AGN of the XBS survey (filled points) and the elusive
AGN (stars). The [OIII] luminosities are corrected only for the Galactic 
extinction.  
The objects with a
``passive'' optical spectrum (i.e. the XBONG-like) are marked with a circle.
On the same plot we report the line corresponding to 
L$_{[2-10 keV]}$/L$_{[OIII]}$=1000 that has been recently proposed 
to classify more properly the XBONG and to separate them from the rest of 
the AGN (Cocchia et al. 2007). 

Fig.~\ref{lumlum}  shows the (expected) correlation between 
the two luminosities but it does not reveal a different behaviour of the
elusive AGN in respect to ``confirmed'' type~1 AGN: 
In the low-luminosity regime (L$_X<$10$^{43}$ erg s$^{-1}$) the expected
luminosity of the [OIII]]$\lambda$5007\AA\ emission line is simply too
low to be detected out from the host-galaxy light. 

\begin{figure}
   \centering
    \includegraphics[width=9cm]{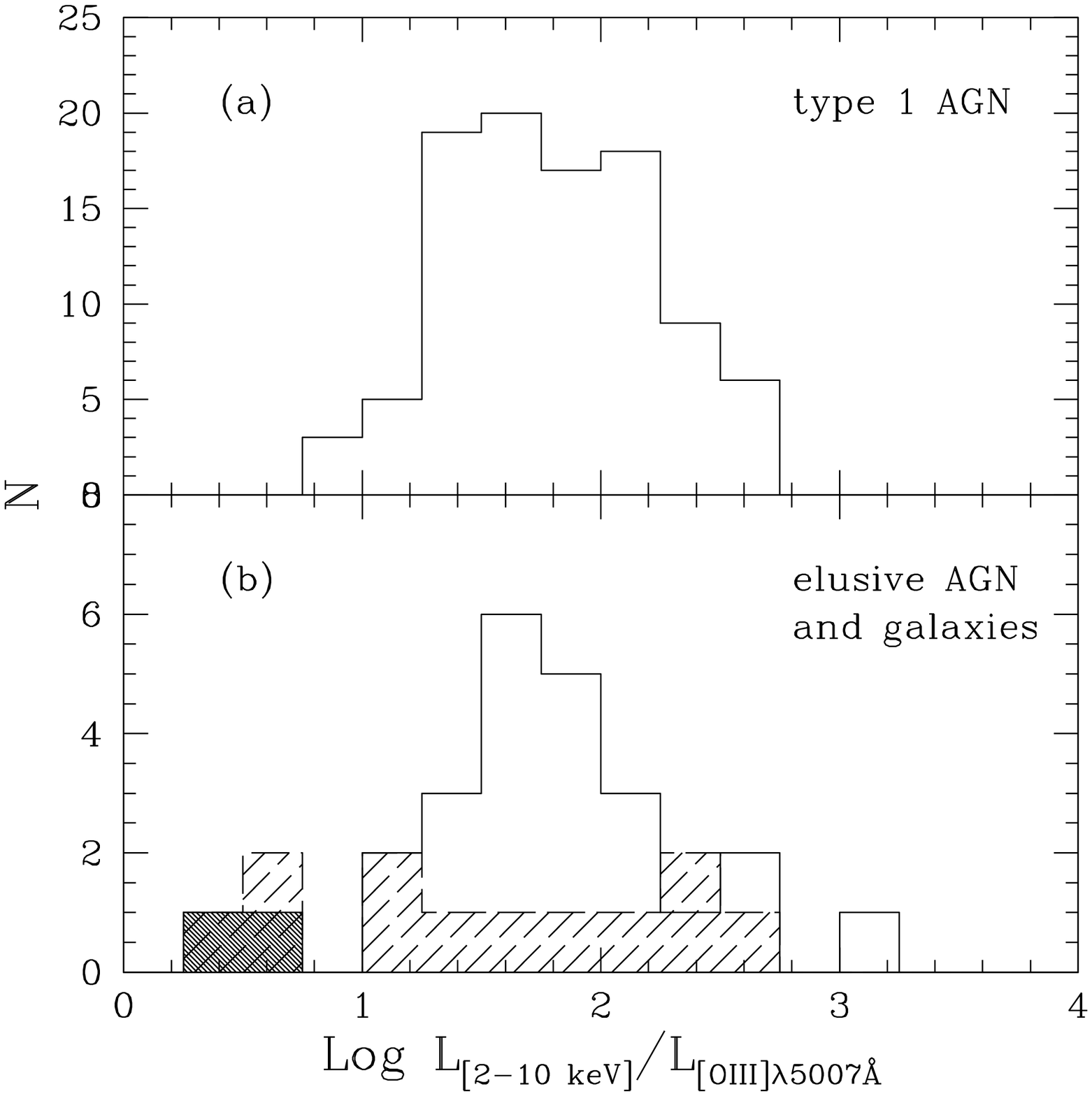}

   \caption{The distribution of the 2-10 keV/[OIII]$\lambda$5007\AA\ 
luminosity ratio for the type~1 AGN (panel a) and the elusive AGN plus
normal galaxies (panel b). Shaded histogram represent the lower limits for the sources
where the [OIII] emission line is not detected. The two sources classified
as normal galaxies are marked with an heavier shadowing.  
}
              \label{lumrat}
    \end{figure}

This result is even more evident by plotting the  distribution of the 
L$_{[2-10keV]}$/L$_{[OIII]}$ ratios for the type~1 AGN and the elusive AGN
(Figure~\ref{lumrat}). The majority of the elusive AGN have 
L$_{[2-10keV]}$/L$_{[OIII]}$ ratios well within the distribution 
observed in the type~1 AGN. 
In the XBS survey we have only one object 
(XBSJ043448.3$-$775329) which
is located in the region below the L$_{[2-10 keV]}$/L$_{[OIII]}$=1000 
line plus one additional source (XBSJ052128.9$-$253032) whose upper limit
on the [OIII]$\lambda$5007\AA\ luminosity is just above this line.

%                                     One column figure (place early!)
%______________________________________________ Gamma_1 (lg rho, lg e)

%                                     One column figure (place early!)
%______________________________________________ Gamma_1 (lg rho, lg e)

We have analysed in more detail these 2 objects. 
Starting from the results of the X-ray analysis (L$_X$ and N$_H$) 
we have applied the spectral model as discussed in Severgnini et al. (2003) 
to constrain the value of the nuclear $\alpha_{OX}$. 
In summary, given the X-ray luminosity and an assumption on the
value of $\alpha_{OX}$ we compute the intrinsic optical luminosity
of the nucleus (before the absorption) and consequently 
normalize the AGN template. We then absorb the continuum and the broad lines
of the template according to the value of N$_H$ found from the X-ray data 
and sum the result to the host-galaxy template. The normalization of the
galaxy template is fixed by reproducing the continuum of the 
observed optical spectrum. The nuclear $\alpha_{OX}$ is then estimated 
as the value that best reproduces the observed spectrum. 
In the case of the 2 objects discussed here
the values of $\alpha_{OX}$ inferred from this analysis are $\leq$1.1,
for XBSJ052128.9$-$253032 while we
find $\alpha_{OX}\sim$1.1 for XBSJ043448.3$-$775329. The two values are 
extreme but still within the distribution
of $\alpha_{OX}$ values observed in  type~1 AGN in the XBS survey 
(that have $\alpha_{OX}$ between 1 and 1.8, with a mean value of
$\alpha_{OX}$$\sim$1.35).

In conclusion, there are no strong indications that the elusive AGNs 
selected in the XBS survey  have values of  $\alpha_{OX}$ or of  
L$_{[OIII]}$/L$_X$ ratio
significantly different from those observed in the entire
AGN population selected in the survey  once the intrinsic spread on the $\alpha_{OX}$ 
or the   L$_{[OIII]}$/L$_X$ ratio
is considered.

\subsection{BL Lac objects}
At least one elusive AGN, selected in the {\it Hellas2XMM} survey, has been classified
as a possible BL Lac nucleus (Brusa et al. 2003). Indeed, the recognition problem
is particularly critical for BL Lac objects due to their characteristic featureless
continuum (Browne \& Marcha 1993) and the fraction of mis-classified BL Lacs
in medium/deep X-ray survey could be significant (March\~a \& Browne 1995). 
However, given the relative 
low space density of this kind of AGN 
it is unlikely that the BL Lac hypothesis can account
for more than ``a few'' elusive AGN. 
Indeed, on the basis of the X-ray logN-LogS of BL Lacs derived from the 
{\it EMSS} survey 
(Wolter et al. 1991), at the flux limit of the XBS survey we  expect 
about 2-7 BL Lac objects in total, a range which is consistent with the number of BL Lac
(or candidates) already detected in the survey (5, Caccianiga et al. 2007, in prep.). 
Therefore,
we do not expect that many (1-2 at most) BL Lac have been mis-classified in the XBS survey
and included among the ``elusive AGN''.

Since BL Lac objects
are radio emitters, it is possible to use the radio detection as a method to find out
the mis-classified BL Lacs out of the 33 elusive AGN in the XBS survey. No dedicated
radio follow-up has been scheduled for the XBS objects. However we can use the existing
radio surveys, both in the north (NRAO VLA Sky Survey, NVSS, Condon et al. 1998) and in 
the south (SUMSS, Bock, Large \& Sadler 1999; Mauch et al. 2003), to search
for a radio detection. 
Out of the 33 elusive
AGN in the XBS survey only two have been detected in the radio band: one 
(XBSJ012654.3+191246) has been 
detected in the NVSS (integrated flux of 1408.1 mJy and peak flux of 947.5 at 1.4 GHz)
and the second (XBSJ231546.5-590313) has been detected in the SUMSS (37.3 mJy at 843~MHz). 
The first source is a radio-galaxy, i.e. it
belongs to the so-called ``parent-population'' of BL Lac objects but the relativistic jets
are not aligned with the observer (Galbiati et al. 2005). 
The second source is the ULIRG IRAS~23128$-$5919, previously discussed, whose radio
emission is likely due to the intense star-formation. 

In conclusion 
we have not found any  good BL Lac candidate among elusive AGN of the XBS survey. 
In principle deeper radio follow-up may detect some additional 
BL Lac candidate, if present. 
However, unless the actual cosmological properties (e.g. luminosity function, cosmological 
evolution and LogN-LogS) of BL Lacs are significantly
different from those estimated until now, we do not expect more than $\sim$2 BL Lacs 
among the elusive AGN of the XBS survey. Therefore, it is unlikely that the BL Lac hypothesis 
can  offer 
a viable explanation for the majority of the elusive AGN in the XBS survey.

\section{Summary and conclusions}
We have studied in detail the 35 objects in the XBS survey whose
optical spectrum is dominated by the light from the host-galaxy and it 
does not reveal the presence of an AGN (20 objects),
or its presence is only suggested (15 objects). In the former group we have
both objects showing an early-type galaxy spectrum (11 objects), and
objects with an optical spectrum characterized by narrow emission lines
with relative intensities suggesting star-formation and/or a LINER
classification  (9 objects). 
The latter group includes sources where  the presence of an
AGN is suggested by some indicators like a possibly broad (FWHM$>$1000 km s$^{-1}$)
H$\alpha$ (10 objects) or a (relatively) strong [OIII]$\lambda$5007\AA\ and no
H$\beta$ detected (5 objects).

We have used the X-ray spectral analysis, based on {\it XMM-Newton} epic 
data 1) to detect the AGN (if present); 2) to classify the sources into
absorbed (N$_H>$4$\times$10$^{21}$ cm$^{-2}$) and unabsorbed (or weakly
absorbed, N$_H<$4$\times$10$^{21}$ cm$^{-2}$) AGNs. The limit on N$_H$ chosen
to discriminate between absorbed and unabsorbed AGN has been fixed in order 
to match 
the optical classification into type~2 and type~1 AGN adopted in the XBS 
survey (Caccianiga et al. 2007, in prep.), that corresponds to A$_V\sim$2 mag.
We have then studied the origin of the optical ``dullness'' in those sources
in which the presence has been assessed by the X-ray analysis.
We summarize the main conclusions.
\begin{itemize}

\item
The X-ray spectral analysis of the 35 objects has indicated the presence
of an AGN in 33 out of 35 sources. This means that, at the flux limit of the 
XBS survey ($\sim$10$^{-13}$ erg s$^{-1}$ cm$^{-2}$)
11\% of the AGN is highly contaminated by the star-light from the host galaxy
(elusive AGN) at a level that the correct optical classification of the
source (or even the detection of the AGN) is difficult or impossible.;

\item
The importance of this ``recognition problem'' is largely dependent on the
X-ray luminosity (Figure~\ref{frac}) and it increases going from 
X-ray luminous objects (L$_X>$10$^{44}$ erg s$^{-1}$), only slightly affected
by this problem (1.5\%), to objects with ``intermediate'' 
X-ray luminosity (10$^{43}$-10$^{44}$ erg s$^{-1}$), where the percentage
of elusive sources becomes significant ($\sim$14\%) and to 
sources with low X-ray luminosity ($<$10$^{43}$ erg s$^{-1}$), where
the recognition problem affects more than 60\% of the sources. This result
suggests that the recognition problem is mainly due to the 
global weakness of the source rather than to a particular optical
``dullness'' (in respect to the X-ray luminosity) of the AGN at least for 
X-ray luminosities below 
$\sim$10$^{43}$ erg s$^{-1}$. For luminosities $>$10$^{43}$ erg s$^{-1}$
also the absorption plays an important role which becomes dominant 
for luminosities above 10$^{44}$ erg s$^{-1}$;

\item
The recognition problem  affects most, in percentage, type~2 AGN 
($\sim$40\% are elusive) while it is less important for type~1 AGN 
(8\% are elusive). Given the larger number of type~1 AGN
in our samples (even in the HBSS sample), however, most ($\sim$57\%) of the elusive
AGN in the XBS survey contain a type~1 AGN. Even if we exclude from the
analysis those elusive AGN for which some hint for the presence of an AGN
can be inferred from the optical spectrum (e.g. those sources with a possibly
broad H$\alpha$ and the sources with a strong [OIII]$\lambda$5007\AA\ emission
line) the number of type~1 AGN among the elusive AGN remains significant
(50\%); 

\item 
We have tested the possibility that a Compton-thick AGN is hiding in some of 
the elusive sources of the XBS survey. To this end we have computed the 
Compton-thickness parameter, T, defined as the ratio of the X-ray and the
[OIII]$\lambda$5007\AA\ flux. No evident Compton-thick candidates have been
found although we  cannot completely rule out this hypothesis for two sources;

\item
We have used a simple spectral model, composed by an AGN and host-galaxy 
template, to better study the elusive AGNs;
Using this model, we have shown
that for values of AGN/galaxy luminosity  ratios (measured at 4050\AA\ and 
through the spectrograph slit) 
above 3, the AGN can be easily recognized
in the optical in the large majority of cases (95\%) while, for values below 3, the
problem of star-light dilution makes the optical AGN detection and/or classification
difficult or impossible for a large fraction (55\%) of the sources;

\item
By studying the relation between the X-ray luminosity and the 
[OIII]$\lambda$5007\AA\ luminosity (which is an indicator of the intrinsic 
optical luminosity of the AGN) we find that the elusive AGN  do not
show evidence for L$_X$/L$_{[OIII]}$ ratios or 
X-ray-to-optical spectral indices ($\alpha_{OX}$)  significantly 
different from those observed in non-elusive AGN. Therefore, at the
relatively brigth X-ray fluxes sampled by the XBS survey we do not
find the ``extreme'' types of XBONG selected at fainter fluxes by Cocchia et al. (2007)
showing high values of  L$_X$/L$_{[OIII]}$ ratios.

\item 
Using the available radio data we have not found any BL Lac candidate among
the 33 elusive AGN. Given the X-ray flux limit of the sample and its sky coverage  
and using the current knowledge on the BL Lac sky-density, 
we conclude that at most 2 elusive AGN may turn out to be low-luminosity
BL Lac objects. The predicted number may be different if the actual statistical properties of 
BL Lacs will turn out to be significantly different from our current best estimates.
  
\end{itemize}

Based on these results, we conclude that the main reason why an important 
fraction of the AGN appears elusive in the XBS survey 
is the global strength of the AGN, coupled in some cases with the
presence of absorption, and it is not necessary to invoke the explanations 
(``non-standard'' AGN or an additional extra-nuclear absorption
within the host-galaxy) that have been proposed to account for objects (typically
XBONGs) selected at fainter X-ray fluxes.

We finally stress that the increasing importance of the AGN ``recognition problem''
with the decreasing of the X-ray luminosity is a factor
that must be taken into account when deriving the statistical 
properties of the AGN (e.g. the luminosity function, the cosmological
evolution, ...) as a class and, more importantly, when divided into
the subclasses of (optically) absorbed and unabsorbed AGN. This is particularly
critical for deep X-ray surveys. 

\begin{acknowledgements}
We acknowledge helpful discussions with S. Andreon, F. Carrera, F. Cocchia, M.J. March\~a and 
G. Trinchieri. We thank the referee for useful suggestions and a quick review.
Based on observations made with: ESO Telescopes at the La Silla Observatories 
under programme IDs: 070.A-0216, 074.A-0024, 076.A-0267;  
the Italian Telescopio Nazionale Galileo (TNG) operated on the island of La Palma by 
the Fundaci\'on Galileo Galilei of the INAF (Istituto Nazionale di Astrofisica) at the 
Spanish Observatorio del Roque de los Muchachos of the Instituto de Astrofisica de 
Canarias; the German-Spanish
Astronomical Center, Calar Alto (operated jointly by Max-Planck
Institut  f\"{u}r Astronomie and Instututo de Astrofisica de
Andalucia, CSIC). AC, RDC, TM and PS acknowledge financial support from the Italian Space
Agency (ASI), the Ministero dell'Universita´ e della Ricerca (MUR) over the last years.
This research has made use of the NASA/IPAC Extragalactic Database
(NED) which is operated by the Jet Propulsion Laboratory,
California Institute of Technology, under contract with the
National Aeronautics and Space Administration. The research described in this 
paper has been conducted within  the {\it XMM-Newton Survey Science Center} 
(SSC, see http://xmmssc-www.star.le.ac.uk.) collaboration, involving a consortium of 
10 institutions, appointed 
by ESA to help the SOC in developing the software analysis system, 
to pipeline process all the {\it XMM-Newton} data, and to exploit the 
{\it XMM-Newton} serendipitous detections. 
\end{acknowledgements}

\appendix

\section{The reliability of the fluxes derived from the optical spectra}
The optical spectra of the XBS sources have been 
taken with the primary aim of deriving the spectral classification and the redshift.
Since both quantities do not require a reliable absolute calibration the observing runs
are not necessarily carried out under photometric conditions. In spite of this, for the
elusive AGN  described in this paper, the [OIII]$\lambda$5007\AA\ flux 
gives important pieces of information to assess the nature of these objects.
In this section we briefly
explain how we have estimated the reliability of the fluxes derived from the optical
spectra and how we have implemented statistical corrections to minimize the (systematic)
errors on the absolute calibration of the spectra.

%                                     One column figure (place early!)
%______________________________________________ Gamma_1 (lg rho, lg e)
   \begin{figure}
   \centering
    \includegraphics[width=9cm]{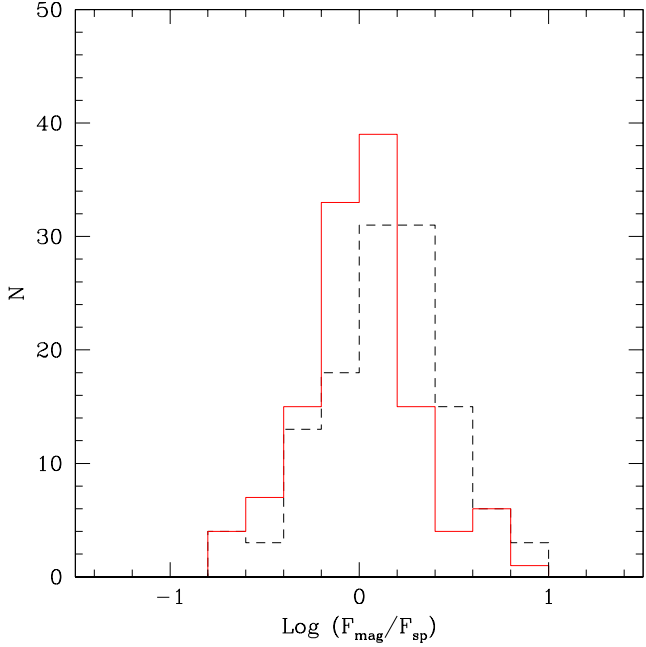}

   \caption{The distribution of the ratio between the fluxes at 6400\AA\  
estimated from the R magnitude and the ones computed from the optical
spectra of the point-like objects before (dotted line) and after (continuous line) the systematic
corrections applied to the sources in each observing run.}
              \label{hispoint}
    \end{figure}
Several facts may affect the absolute calibration of an optical spectrum. The most important ones
are 1) the sky transparency when the objects and the standard star are observed, 2) the seeing conditions
with respect to the slit width used during the night, 3) the flux loss across the slit for extended targets. 
A worsening of the sky transparency and/or the seeing produces a systematic flux loss if it happens 
during the observation of the source while it produces a systematic increment of the computed 
fluxes if it happens during the observation of the standard star. The 
flux loss across the slit in extended targets only affects the host galaxy emission and not 
the nuclear emission like, for instance, the nuclear [OIII]$\lambda$5007\AA\ flux. The ``a posteriori'' 
correction of all these parameters is not trivial. A possible solution could be a re-normalization of each
spectrum with the photometric points obtained independently. The problems connected 
with this procedure are: 1) the magnitudes available for the XBS sources are typically taken 
from APM and are affected by a large uncertainty ($\approx$0.5 mag) and 2) the elusive sources are, 
by definition, extended sources and, thus, in order to re-normalize the spectrum to the photometric point 
we need to estimate the fraction of light from the galaxy lost because of the limited size of slit. 
Given these two main difficulties we decided to adopt a statistical approach using the point-like
objects (for which the latter effect is not important) and comparing the fluxes measured from the spectra to the 
fluxes  estimated from the magnitudes. In particular we compute 
the global spread and any
possible systematic shifts using the point-like sources in each of the observing run. We then apply
a systematic correction to the spectra of those runs where a significant shift is measured and compute
the global spread of the final distribution. 
In Figure~\ref{hispoint} we show the results of this analysis i.e. the histogram 
of the ratio between the 2 fluxes (computed from the spectra and estimated from the R magnitude) before and after
the systematic corrections on each observing run. The final spread, after the correction is $\sigma$=0.30.
This spread includes both the uncertainties on the spectral fluxes and the errors on the magnitudes 
($\sigma_{mag}\sim0.2$) used for the
analysis. The intrinsic spread on the spectral fluxes is then given by:

\begin{center}
$\sigma_{Log flux} = \sqrt{\sigma^2 - \sigma_{mag}^2} \sim 0.22$
\end{center}

Which corresponds to a relative  error on the fluxes of about 50\%.
The points plotted in all the Figures including the [OIII]$\lambda$5007\AA\ 
fluxes or luminosities  
are corrected with the systematic shifts discussed above and
the reported error bars correspond to the value of $\sigma_{Log flux}$.

\section{Notes on individual sources}

{\bf XBSJ100032.5+553626}. The optical classification of this object is controversal in the
literature: from the one hand it has been classified as Seyfert~2 by Mason et al. (2000) 
and, on the other hand, it has been included in a list of $\sim$2000 NLSy1 extracted from the SDSS 
by Zhou et al. (2006).
The SDSS optical spectrum shows strong emission lines on a blue continuum
(the 4000\AA\ break is =7$\pm$5\%). 
The widths of the permitted lines (e.g. the H$\beta$) are
very small ($<$400 km/s) and much narrower than the narrowest permitted lines observed in the NLSy1 of the XBS survey
(the  H$\beta$ widths of the NLSy1 in the XBS survey range from 900 to 1800 km/s, Caccianiga et al. 2007, in prep.). Additionally 
the strenght of the FeII$\lambda$4570\AA\ normalized to the H$\beta$ line (the R$_{4570}$ parameter) is 
lower (0.28, Zhou et al. 2006) than what is usually observed in NLSy1 (typically  R$_{4570}>0.50$,  
e.g. V\'eron-Cetty et al.  2001). Finally, the images taken from the SDSS clearly show an extended
source without obvious indication of a strong nucleus. For all these reasons we do not adopt the NLSy1
classification and we prefer a classification as starburst. Indeed, the critical line ratios 
([OIII]$\lambda$5007\AA/H$\beta$ and [NII]$\lambda$6583/H$\alpha$) put the source in the starburst/AGN+starburst
composite spectrum region in the Veilleux \& Osterbrock (1987) diagnostic plot.

\noindent
{\bf XBSJ143911.2+640526 }. This source has a ``passive'' optical spectrum with a relatively
low 4000\AA\ break (33$\pm$7\%). The lack of emission lines coupled with the reduced break 
is suggestive of the presence of a significant featureless nuclear emission. 
Therefore  we first tentatively
classified of this source as a BL Lac object candidate (Della Ceca et al. 2004). 
However the shape of the optical spectrum
is not very well determined due to the bad observing conditions (cloudy night) and the lack of
a standard star observed during the same night. Moreover the observed value of 4000\AA\ can still
be consistent with those observed in elliptical galaxies. 
Therefore we consider this object as ``elusive'' 
AGN candidate and used the X-ray analysis to find and characterize the hidden AGN. The X-ray spectral
analysis has revealed the presence of an absorbed AGN (N$_H$=2$\times$10$^{23}$ cm$^{-2}$). The X-ray
spectrum is not consistent with the BL Lac hypothesis due to the large column density. These sources,
in fact, usually show 0.5-10 keV spectra characterized by unabsorbed or slightly (a few 10$^{21}$ 
cm$^{-2}$) 
absorbed PL (e.g. Perlman et al. 2005) which can be alternatively fit with a model with an intrinsic 
curvature. Additionally, 
this source has not been detected in the radio (in the NVSS survey) something that can be
considered as a further indication (although not conclusive, see discussion in Section~4) 
against the BL Lac hypothesis. We therefore classify the source as type~2 AGN.

\end{document}